\documentclass{aa}  

\usepackage{graphicx}
\usepackage{txfonts}
\usepackage{lipsum}
\usepackage{subcaption}         
\usepackage{lscape}             
\usepackage{placeins}           
                                
\usepackage[colorlinks=true,allcolors=blue]{hyperref}

\newcommand{\fesc}{$f_{\rm{esc}}$\xspace}

\newcommand{\ha}{H$\alpha$}

\begin{document}

   \title{Reionization driven by the few: the ionizing budget of galaxies at z=5-10 from JWST/NIRSpec}\titlerunning{Ionizing emissivity in the EoR}

%

\author{Emma Giovinazzo\inst{1}\thanks{E-mail: emma.giovinazzo@unige.ch}
\and
Pascal A. Oesch\inst{1, 2, 3}
\and
Anne Verhamme\inst{1}
\and
Romain A. Meyer\inst{1}
\and
Callum Witten\inst{1}
\and
John Chisholm\inst{4}
\and 
Rui Marques-Chaves\inst{1}
\and
Charlotte Simmonds\inst{5}
        }

   \institute{
Department of Astronomy, University of Geneva, Chemin Pegasi 51, 1290 Versoix, Switzerland
\and 
Cosmic Dawn Center (DAWN), Niels Bohr Institute, University of Copenhagen, Jagtvej 128, K\o benhavn N, DK-2200, Denmark
\and
Niels Bohr Institute, University of Copenhagen, Jagtvej 128, Copenhagen, Denmark
\and
Department of Astronomy, The University of Texas at Austin, Austin, TX 78712, USA
\and
Departamento de Astronomía, Universidad de Chile, Camino El Observatorio 1515, Las Condes, Santiago}

   \date{Received XXX}

 
  \abstract
{The sources responsible for the Epoch of Reionization (EoR), the last major phase transition of the Universe, remain highly elusive. While JWST has begun to illuminate the properties of early galaxies, the direct detection of their ionizing photons is virtually impossible due to the neutral inter-galactic medium (IGM) at $z\gtrsim 6$.
A direct estimate of the escape fraction of ionizing photons $f_{\rm esc}$ is thus not possible. However, the escape fraction is encoded in spectral features that trace a low opacity to ionizing photons, namely the reduction of nebular emission with increasing $f_{\rm esc}$.
Here, we exploit the large archive of JWST/NIRSpec spectra at $5< z_{\rm spec}<10$ from the DAWN JWST Archive to provide new constraints on the ionizing emissivity of EoR galaxies and on the timeline of reionization.
Through detailed spectral fitting of 1428 galaxies with a picket-fence model for the escape of ionizing photons, we derive posteriors of \fesc\ together with the effective ionizing output. This approach yields the escaping ionizing output of each galaxy while bypassing the usual decomposition into the ionizing photon production efficiency ($\xi_{\rm ion}$) and \fesc. Combining these measurements with UV luminosity functions (LFs), we derive the hydrogen ionizing LFs, resulting in an observationally-derived ionizing photon budget at $5< z <15$, integrated down to $\rm M_{UV}=-13$.
Under standard assumptions, our budget yields an evolution of the IGM neutral fraction consistent with independent probes, with reionization ending at $z\sim5.8$. Given the evolving shape of the ionizing LF, bright sources ($\rm M_{UV}\sim-20$) dominate the cosmic ionizing emissivity $\dot{n}_{\rm ion}$ at lower redshift ($z=5-6$), whereas at $z>6$ faint galaxies ($\rm M_{UV}>-18$) contribute roughly equally. Remarkably, a mere $\sim$20\% of sources, those with $f_{\rm esc}>10\%$, produce $\sim87\%$ of the ionizing photons.
Overall, our results support a picture in which a few strongly leaking galaxies drive most of reionization.}


   \maketitle
   \nolinenumbers

\section{Introduction}

The epoch of reionization (EoR) is the last major phase transition of the Universe, when the hydrogen in the intergalactic medium (IGM) transitioned from completely neutral to completely ionized \citep[][for some recent reviews]{Becker_15, Dayal_18, Robertson_22, Fan_23, Ellis_25, Stark_26}, finally allowing Lyman Continuum (LyC) photons ($\lambda_{\rm rest}<912$ \AA) to travel through it mostly unimpeded. This process started with the formation of the first luminous objects \citep{Barkana_Loeb_01} and was likely driven by star-forming galaxies \citep[e.g.][]{Bouwens_15, Robertson_15, Dayal_25}. However, there is still debate about whether faint \citep{Atek_24} or bright \citep{Naidu_20} galaxies dominate this transition. Active Galactic Nuclei (AGNs) have also been suggested as important contributors to this process \citep{Giallongo_15, Madau_15, Grazian_24, Madau_24}, although studies on their luminosity function in the EoR suggest that their contribution is likely minor, compared to that of galaxies \citep{Mitra_18, Hassan_18, Kulkarni_19, Jiang_22, Matsuoka_23}. 

In this context, the process of reionization is expected to be very patchy \citep[e.g.,][]{DAloisio_15, Jamieson_25}, with bubbles of ionized gas forming and growing around the objects that leak the most ionizing photons. This is seen directly in observations of the Gunn-Peterson trough, which show field-to-field variations towards the end of the EoR \citep{Bosman_18, Eilers_18, Yang_20, Bosman_22, Meyer_25}. Although inhomogeneous, the global evolution of reionization can be summarized with the volume filling fraction of ionized hydrogen $Q_{\rm HII}(z)$ and its counterpart, the fraction of neutral hydrogen $x_{\rm HI}$. 
There have been significant efforts to constrain $x_{\rm HI}$ and its redshift evolution, both from single bright objects that can probe the Gunn-Peterson trough, such as quasars, and from the redshift evolution of galaxy populations. Constraints on $x_{\rm HI}$ have been placed from observations of the Lyman-$\alpha$ and Lyman-$\beta$ absorption in the spectra of quasars \citep{Eilers_18, Kulkarni_19, Yang_20, Bosman_22, Qin_24} and galaxies \citep{Meyer_25}, the clustering of Lyman-$\alpha$ emitters \citep{Sobacchi_15, Ouchi_18}, the Lyman-$\alpha$ equivalent width and luminosity function evolution \citep{Ouchi_10, Konno_14, Konno_18, Zhengh_17, Inoue_18,Mason_18, Mason_19b, Hoag_19, Whitler_20, Goto_21, Morales_21, Bolan_22, Ning_22, Bruton_23, Morishita_23, Nakane_24, Tang_24, Kaguera_25} as well as the damping wing of either Lyman-break galaxies \citep{Hsiao_24, Umeda_24, Mason_26} or quasars \citep{Schroeder_13, Davies_18, Wang_20, Greig_19, Greig_24}. All these different estimates place the end of the reionization process at $z\sim5.3-6$.

In its most basic form, the redshift evolution of the fraction of neutral hydrogen is calculated as the balance between the recombination time scale and the rate of ionization \citep{Madau_99}. The number of ionizing photons that reach the IGM per unit time and unit volume, the cosmic ionizing emissivity, is described as 
\begin{equation}\label{eq:nion}
    \dot{n}_{\rm ion} = \rho_{\rm UV}f_{\rm esc}\xi_{\rm ion}.
\end{equation}
Here $\rho_{\rm UV}$ is the UV luminosity density (in units $\rm erg \, s^{-1} \, Hz^{-1} \, Mpc^{-3}$), derived from the UV luminosity function, $f_{\rm esc}$ is the average escape fraction of ionizing photons of the general galaxy population and $\xi_{\rm ion}$ is the average ionizing photon production efficiency (in units $\rm Hz \, erg^{-1}$), measuring the amount of ionizing photons created per 1500 \AA\ luminosity density. 

Since the launch of the James Webb Space Telescope \citep[JWST,][]{Gardner_23} we have gained an unprecedented view of the rest-frame optical properties of galaxies in the EoR, both thanks to the Near-Infrared Camera  \citep[NIRCam,][]{Rieke_23} and the Near-Infrared Spectrograph \citep[NIRSpec][]{Jakobsen_22}. In particular, many studies have focused on separately measuring the three parameters that make up our understanding of the ionizing photon output of the Universe, $\rho_{\rm UV}$ \citep{Bouwens_23, Donnan_24, Harikane_25, Whitler_25, Atek_26, Chemerynska_26, Kreilgaard_26, Weibel_26}, $\xi_{\rm ion}$ \citep{Simmonds_24a, Simmonds_24b, Llerena_25, Pahl_26}, and $f_{\rm esc}$ \citep{Witten2023, Mascia_24, Saxena_24, Giovinazzo_25, Stoffers_26, Jecmen_26}. However, considering each component individually does not account for possible correlations between the parameters, which can have significant implications on the estimated ionizing budget \citep{Munoz_24, Bosman_24}. In particular $f_{\rm esc}$ is an exceedingly challenging quantity to measure as effectively all ionizing photons are absorbed by the IGM in the EoR, making direct measurements impossible. A complete analysis where all the parameters are modeled together, self-consistently, is therefore necessary to fully understand the reionization process.


\cite{Giovinazzo_25} estimated \fesc for a sample of sources with NIRSpec/PRISM spectroscopy in the EoR, from Spectral Energy Distribution (SED) fitting. In this work, we will expand on this analysis to estimate the rate of ionizing photon production ($\dot{N}_{\rm ion, int}$) for the same sample, again using SED-modelling. As a result of the \fesc prescription already implemented by \cite{Giovinazzo_25}, we can infer posterior distributions for the ionizing photon emission rate ($\dot{N}_{\rm ion, esc}$). Similar efforts in estimating the ionizing emissivity from SED fitting have also been undertaken by \cite{Simmonds_24b}, with a mass complete sample, although without accounting for \fesc. 
By estimating the ionizing photon emission rate directly through integration of the ionizing part of the spectral SED fits, we avoid having to disentangle the separate contributions of the production efficiency and the escape fraction, along with their associated uncertainties, to $\dot{N}_{\rm ion}$. This yields a self-consistent estimate of the rate at which ionizing photons reach the IGM from a given galaxy, a quantity necessary to constrain the reionization process.

In this paper, we will determine the evolution of the ionizing emissivity function of the general galaxy population as well as the cosmic ionizing emissivity. We will then calculate the redshift evolution of the neutral hydrogen fraction with cosmic time, as well as the contribution of strong leakers, here defined as \fesc$> 10\%$. We will test which population of galaxies contributes most to the ionizing emissivity, whether the few highly leaking galaxies or the majority of weakly leaking sources. We will also determine whether any galaxy population, i.e. at faint, intermediate or bright luminosities, is a dominant contributor of ionizing photons to the IGM.

In Sec.~\ref{Method} we introduce the sources included in our sample, present the method that we used to calculate $\dot{N}_{\rm ion}$ and compare it to the method typically used in the literature. In Sec.~\ref{Ionizing_properties} we present the ionizing properties of the sources in our sample and their redshift and magnitude evolution. In Sec.~\ref{LF} we present the ionizing emissivity function, quantifying the number of sources in a unit volume with a given rate of ionizing photon output. In Sec.~\ref{Neutral_fraction} we present the evolution of the fraction of neutral hydrogen. Finally, in Sec.~\ref{Discussion} and Sec.~\ref{Conclusion} we present our caveats and conclusions.

Throughout this work, we assume a flat $\Lambda$CDM cosmology with $H_0=70\,\textrm{km}\,\textrm{s}^{-1}\,\textrm{Mpc}^{-1}$, $\Omega_m=0.3$, and $\Omega_\Lambda=0.7$. Magnitudes are given in the AB system \citep{Oke_Gunn_83}.



\section{Ionizing output calculation}\label{Method}

\subsection{Spectroscopic data}

For this study, we used the data set previously presented in \cite{Giovinazzo_25}. This consists of 1428 sources with JWST NIRSpec/PRISM observations at $5<z_{\rm spec}<10$ taken from the DAWN \textit{JWST} Archive (DJA\footnote{\url{https://dawn-cph.github.io/dja/}}) \citep{Heintz_25, brammer_2025}, an online repository containing spectroscopic data from public \textit{JWST} programs, all uniformly extracted and reduced with the same pipeline based on \texttt{grizli}\footnote{\url{https://github.com/gbrammer/grizli}} and \texttt{MSAexp}\footnote{\url{https://github.com/gbrammer/msaexp}}. Further details on the data reduction and processing can be found in \citet{deGraaff_24,Heintz_25,Valentino25,Pollock25}. The database query was performed on the 7th of February 2025 and includes all sources in the redshift range of interest with grade=3 (i.e., with robust redshift), M$_{\rm UV}<-18$ and excludes known Little Red Dots (LRDs) as reported by \cite{Kocevski_24}. 
This selection includes spectra from RUBIES (GO 4233, \citealt{deGraaff_24})  JADES (GTO-1180, 1181, 1210, 1286, 1287; GO-3215 \citealt{Eisenstein_23a}) GTO WIDE (GTO-1211, 1212, 1213, 1214, 1215; \citealt{Maseda_24}), CEERS (ERS-1345; \citealt{Finkelstein_23}), UNCOVER (GO-2561, \citealt{Bezanson_24}), CAPERS (GO-6368; \citealt{Donnan_25}), GO-1433 \citep{Hsiao_24}, GO-2198 \citep{Barrufet_25}, GO-2565 \citep{Nanayakkara_24}, DD 2750 \citep{ArrabalHaro_24}, GO-3073 \citep{Castellano_24}, DD-2756 \citep{Mascia_24}, DD-6541 \citep{DeCoursey_25}, DD-6585 (PI: Coulter) and GO-4106 (PI: Nelson).

\subsection{Spectral SED fitting}

For each source we perform spectral SED fitting to the NIRSpec/prism data to recover the physical properties, as described in \cite{Giovinazzo_25}. These fits were performed with \texttt{bagpipes} \citep{Carnall_18, Carnall_19, nautilus} including a picket-fence model \citep{Heckman_01, Zackrisson_13} to model the escape fractions. This formalism for \fesc\ was first introduced as a custom version of \texttt{bagpipes} in \cite{Giovinazzo_25}, but has since been included in the standard release, from version 1.3.3. 

We modeled the star formation history with the \texttt{continuity} prior \citep{Leja_19}, a non-parametric prior, allowing for greater flexibility than parametric models. We modeled the star formation rate (SFR) in seven bins --with 5, 10, 50, 100, 200, 400 and 800 Myr-- unless the age of the Universe at the source's redshift was less than 800 Myr. In that case, the last bin ended at the age of the Universe. With the continuity prior, \texttt{bagpipes} fits for the $\Delta \rm SFR$ between adjacent bins, which therefore adds a number of free parameters equal to the number of bins minus one. We used a Student-t prior with $\Delta \rm SFR \in [-3, 3]$ and adopted $\nu = 2$ and $\sigma = 0.3$ following \cite{Leja_19}. For \fesc we opted for a logarithmic prior with range $0.001<f_{\rm esc}<1$. This prior was chosen since observations \citep{Kreilgaard_24} as well as simulations \citep{Choustikov_24, Yeh_23} suggest that most sources have little-to-no leakage. This was confirmed in \citet{Giovinazzo_25}. We used the BPASS v2.2.1 stellar-population models \citep{Stanway_Eldridge_18} with the default broken-power-law initial mass function (IMF), with slopes of $\alpha_1 = -1.30$ for the $0.1 - 0.5 \rm M_{\odot}$ mass range and $\alpha_2=-2.35$ for the $0.5 - 300 \rm M_{\odot}$ mass range, based on \cite{Kroupa_93}. Metallicity was modeled with a logarithmic prior, in the range $0.01<Z/Z_{\odot}<0.5$. The stellar and nebular attenuation were modeled with the Calzetti dust law \citep{Calzetti_00}, with a linear prior, allowing for a maximum $A_{V} = 0.5$ mag as we were mainly interested in blue sources. We note that for our picket-fence model dust is not applied to the free channels.  For more details on the SED fits we refer the reader to \cite{Giovinazzo_25}.

The SED fits are used to calculate the ionizing photon production for each source, as discussed in Sec. \ref{Nion_SED}. This is then compared to the more standard approach, presented in Sec. \ref{Nion_ha}, of measuring the ionizing photon production from the H$\alpha$ line.

\subsection{$\dot{N}_{\rm ion}$ estimation from \ha}\label{Nion_ha}

In the standard approach used in the literature, the production rate of ionizing photons in the EoR is estimated from the luminosity of the \ha\ line \citep[e.g.][]{Bouwens_16, Llerena_25, Pahl_26}. As a recombination line, the \ha\ luminosity directly depends on the amount of reprocessed ionizing photons and therefore effectively counts the number of ionizing photons that do not escape. For all the sources in our sample with \ha\ coverage in the PRISM wavelength range, the \ha\ fluxes are extracted following \cite{Covelo_Paz_26}, i.e. by fitting a Gaussian directly to the spectrum. The luminosities are then corrected for dust attenuation following the \cite{Calzetti_00} dust model with attenuation $\rm A_{V}$ derived from SED fitting, as $L_{\rm H\alpha,int} = L_{\rm H\alpha, obs}\times10^{0.4A_{\rm H\alpha}}$.

The relation between the  intrinsic Balmer line luminosity and the total Lyman Continuum luminosity was derived from stellar models by \cite{Leitherer_Heckman_95}. For \ha\ it can be expressed as
\begin{equation} \label{eq:NH0}
    \rm L(H\alpha)[erg \ s^{-1}] = 1.36 \times 10^{-12} N(H^0) [s^{-1}]
\end{equation}
where $\rm N(H^0)$ is the production rate of ionizing photons assuming \fesc = 0, and $\rm L(H\alpha)$ is the intrinsic \ha\ luminosity, which is thus dust-corrected. This effectively assumes that all the ionizing photons are reprocessed in the \ha\ line and none escape. Hence, the ionizing photon production that we find from this method must be corrected for the escape fraction. As seen in \cite{Giovinazzo_25}, most sources have very low escape fraction, making this a good approximation. There is, however, a small fraction of sources with very high escape fraction, as the \fesc\ distribution approximates an exponential with a mean of 10\%. The ionizing photon production rate is then given by
\begin{equation}
    \dot{N}_{\rm ion, int} = \frac{\rm N(H^0)}{1- f_{\rm esc}}.
\end{equation}
For the escape fraction measurements we use those already presented in \cite{Giovinazzo_25}, which are based on spectral SED fitting returning the full posterior distributions.

\subsection{$\dot{N}_{\rm ion}$ estimation from SEDs}\label{Nion_SED}

Alternatively, the ionizing photon production rate can also be calculated as the integral of the spectrum, $L_{\nu}$, below the Lyman limit ($\lambda<$ 912 \AA):
\begin{equation}
    \dot{N}_{\rm ion} = \int^{\infty}_{\nu912} \rm L_{\nu}(\textit{h} \nu)^{-1} d\nu \, , 
\end{equation}
as is done for simulated galaxies \citep[e.g.][]{Naidu_20, Seeyave_23}. We can apply this to calculate both the intrinsic ionizing photon flux, when integrating the intrinsic SED, and the escaping ionizing photon flux, when integrating the observed SED without IGM attenuation. In contrast to what is done with \ha, this method simply counts the photons that escape the galaxy.

Using the full posterior distribution of physical parameters for each source in the sample, we compute the posterior of both the escaping ($\dot{N}_{\rm ion, esc}$) and the intrinsic ionizing photon flux ($\dot{N}_{\rm ion, int}$) for each galaxy. Their respective uncertainties are also computed from the posterior distribution. 
In a first step, we create rest-frame SEDs, to remove the effect of IGM transmission, which still include both the dust and escape fractions to compute $\dot{N}_{\rm ion, esc}$. Then, in a second step, we compute the intrinsic ionizing photon production of the stellar population ($\dot{N}_{\rm ion, int}$) by removing dust obscuration, the effects of IGM transmission and setting $f_{\rm esc}=1$, to create an intrinsic SED.
This results in a posterior distribution of both $ \dot{N}_{\rm ion, esc}$ and $\dot{N}_{\rm ion, int}$ for each galaxy.

\subsection{Comparison of ionizing fluxes}

\begin{figure}
    \centering
    \includegraphics[width=\linewidth]{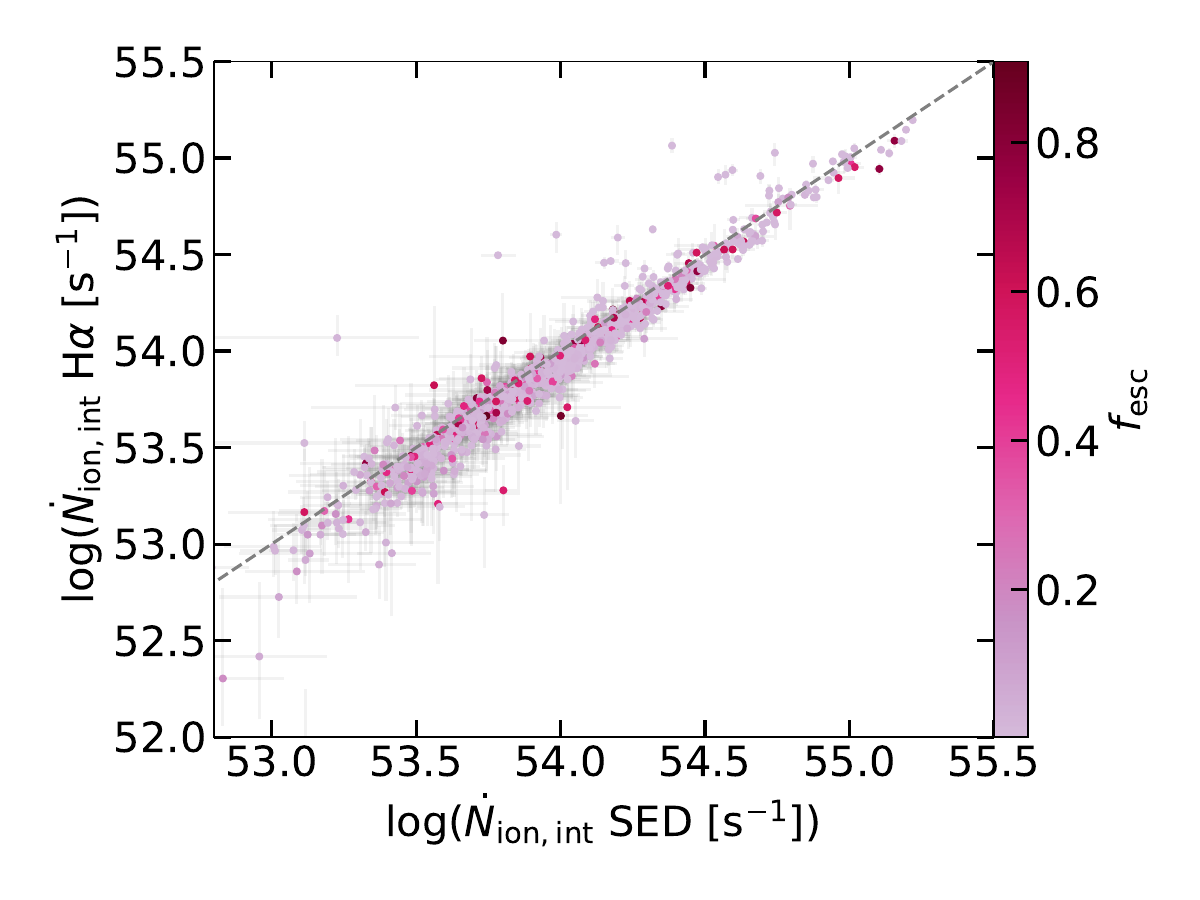}
    \caption{Comparison between intrinsic ionizing photon production rates estimated with \ha\ and from the SED fit, color coded by \fesc. The grey dashed line shows the 1:1 relation. We find a broad agreements, with small offset between the methods ($\Delta = 0.05$ dex), as well as some scatter, with only a few sources lying significantly off the 1:1 line. This agreement with a method widely used in literature \citep[e.g.][]{Bouwens_16, Llerena_25, Pahl_26} validates our SED fitting estimates.}
    \label{fig:method_comparison}
\end{figure}

The two methods described before are complementary in how they count the amount of ionizing photons, as the \ha\ method counts the rate of photons that are reprocessed into nebular continuum, while the SED integration directly counts the rate at which ionizing photons escape. Because of this, the only way to consistently compare these two methods is by using $\dot{N}_{\rm ion, int}$, the intrinsic rate of ionizing photons production. 

In Fig. \ref{fig:method_comparison} we compare the intrinsic ionizing photon production rate as estimated from the two methods. We use this comparison to ensure that our method is consistent with that primarily used in the literature \citep[e.g.,][]{Bouwens_16, Llerena_25, Pahl_26}. 
We find the \ha\ estimates to be consistently lower, with a median difference between the estimates of $\Delta = 0.05$ dex, with only few sources lying significantly above or below the 1:1 line. This small discrepancy is likely due to the assumptions in the $\rm N(H^0)$ conversion to $\rm L(H\alpha)$ of \cite{Leitherer_Heckman_95} that do not completely match those of our \texttt{bagpipes} models. Under the assumption of case B recombination, different gas temperatures and densities between the \texttt{bagpipes} and the \cite{Leitherer_Heckman_95} models could be the source of this discrepancy. Moreover, the two models differ in their IMF. While the high mass slope is the same ($\alpha = -2.35$), the mass range of \cite{Leitherer_Heckman_95} only extends to 100 $\rm M_{\odot}$, unlike the IMF used in BPASS that extends to 300 $\rm M_{\odot}$. Young and massive stars are those expected to produce the most ionizing photons, and it has been shown by \cite{Schaerer_25} that Very Massive Stars (VMS; $\rm M>100 M_{\odot}$) contribute significantly to the Lyman continuum emission of young stellar populations. In general, the stellar population synthesis code in \cite{Leitherer_Heckman_95} is different than that of BPASS, especially in their lack of binaries, which also produce more ionizing photons than single stars. These factors may explain why $\dot{N}_{\rm ion, int}$ as calculated from \ha\ is underestimated when compared to that calculated from our SED fits. This suggests that estimates done with Eq.~\ref{eq:NH0} may not be fully comparable to those that assume a different stellar population synthesis and spectral synthesis codes. 

In Fig. \ref{fig:method_comparison} we also find some scatter with $\dot{N}_{\rm ion, int}$ from \ha\ occasionally being overestimated as compared to estimates from the SED. Visual inspection of these outliers shows that most have a broad H$\alpha$ line, hinting at the possible presence of AGNs or LRDs in these sources. We do not use an AGN template in our fits and therefore find a poor fit to the H$\alpha$ line for these sources.

Overall, we find good agreement between the two methods, indicating a good agreement between our SED fitting results, where \fesc\ is properly accounted for on a source-by-source basis and the method previously used in literature. This validates our model, which we will use for the remainder of the analysis to have access to the ionizing photon production rate also for sources where \ha\ is not covered. In this work, we will mostly use $\dot{N}_{\rm ion, esc}$ as we are interested in the rate at which ionizing photons reach the IGM.


\section{Ionizing properties}\label{Ionizing_properties}

\subsection{Redshift evolution}

\begin{figure} 
    \centering
    \includegraphics[width=\linewidth]{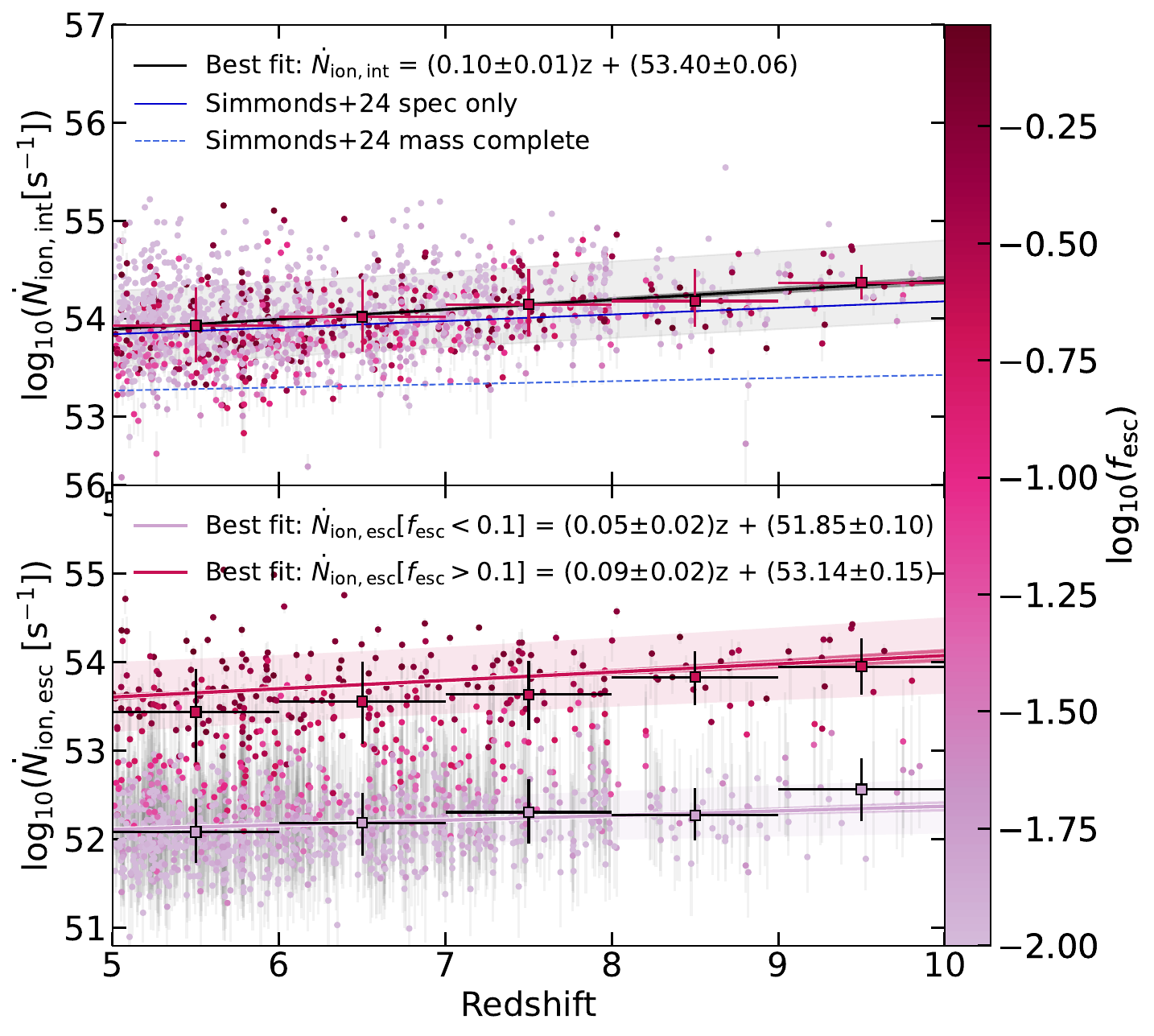}
    \caption{\textbf{Top panel: } Redshift evolution of $\dot{N}_{\rm ion, int}$. The black line shows the best fit for the redshift evolution, with the intrinsic scatter in the gray shaded area. The squares show the binned means. This evolution is consistent with the results from the spectroscopic only dataset from \cite{Simmonds_24b}, shown as the solid blue line, but higher than the one from a mass complete sample, shown in the dashed blue line, also from \cite{Simmonds_24b}. \textbf{Bottom panel:} Redshift evolution of $\dot{N}_{\rm ion, esc}$, color coded by log($f_{\rm esc}$). The pink line with the pink shaded area represents the fit to the $f_{\rm esc}<0.1$ sources and the red line with the red shaded area represents the same for sources with $f_{\rm esc}>0.1$. The two fits sit about 1.5 dex apart, with the high $f_{\rm esc}$ fit being much closer to the intrinsic $\dot{N}_{\rm ion, int}$, as expected. The squares show the binned means, with the same color coding as for the linear fits.}
    \label{fig:ndot_vs_z}
\end{figure}

\cite{Simmonds_24b} have shown that $\dot{N}_{\rm ion}$ evolves with redshift when considering a spectroscopic-only sample, with higher ionizing emissivity at higher redshift. However, they have also demonstrated that this redshift evolution disappears when using a mass complete sample. This suggests a bias of the spectroscopic sample toward star forming galaxies rather than a real change in the ionizing emissivity with redshift for the global galaxy population. An important caveat of this previous analysis is the lack of a prescription for \fesc and therefore the assumption of \fesc = 0 for the calculation of  $\dot{N}_{\rm ion}$. 
We investigate the possibility of a redshift evolution of $\dot{N}_{\rm ion}$, both intrinsic and escaped, in our sample, which includes a prescription for \fesc.

We find that the intrinsic ionizing photon production rate $\dot{N}_{\rm ion, int}$ shows a shallow evolution with redshift, as seen in the top panel of Fig.~\ref{fig:ndot_vs_z}. This trend may be due to a selection bias, as discussed by \cite{Simmonds_24b}. It may also be due to a redshift evolution of $\rm M_{UV}$ or \fesc\  which would translate to a redshift evolution in $\dot{N}_{\rm ion}$, the latter of which would have been missed in previous studies. In \citet{Giovinazzo_25}, we do in fact see a shallow trend between redshift and \fesc for our sources, as also shown in \cite{Ferrara_25}.

In the bottom panel of Fig.~\ref{fig:ndot_vs_z}, we show the rate of escape of ionizing photons $\dot{N}_{\rm ion, esc}$ against redshift. Here we also color-code the points by log$_{10}(f_{\rm esc})$ and see that the data seems to be split in two populations, one clustered around ${\rm log}(\dot{N}_{\rm ion, esc}/ {\rm s^{-1}}) \sim 52$, with consistently low \fesc\ and another one with $f_{\rm esc}\gtrsim10\%$ which spans the range $53<{\rm log}(\dot{N}_{\rm ion, esc}/ {\rm s^{-1}})<55$. It should be noted that the points in the lower cloud, with low \fesc, may be more accurately interpreted as upper limits on $\dot{N}_{\rm ion, esc}$ as the lowest possible value in the fit is \fesc$ = 10^{-3}$, therefore not allowing for a true \fesc = 0. Moreover, as shown in \cite{Giovinazzo_25}, with our spectral SED fitting method it is challenging to confidently constrain escape fractions below $f_{\rm esc}\sim 20\%$ even for high signal to noise values, with \fesc\ usually being underestimated in this case. These uncertainties are properly captured by the posterior distributions, making the uncertainties on $\dot{N}_{\rm ion, esc}$ for sources with low \fesc\ significant. The distribution may therefore not truly be bimodal, as it initially appears. 

We additionally explore the redshift evolution of $\dot{N}_{\rm ion, esc}$, as previously performed for the intrinsic value. We choose to use two linear fits to distinguish the population of strong-leakers ($f_{\rm esc}>10\%$) from that of weak-leakers ($f_{\rm esc}<10\%$). We choose to split the sample at $f_{\rm esc}=10\%$ as this is the mean $f_{\rm esc}$ of the entire sample, as shown in \cite{Giovinazzo_25}. While both populations show a shallow redshift evolution, it is stronger for the high \fesc population, although the evolution is consistent within the uncertainties. This redshift evolution is closer to that of the intrinsic production rate, with a slope of $0.09\pm0.02$, as most ionizing photons that are created are also emitted in the case of high \fesc. Instead, the low \fesc\ population shows a shallower distribution, with a slope of $0.05\pm0.02$, about 1.5 dex below the strong leakers. As mentioned for $\dot{N}_{\rm ion, intr}$ this redshift evolution could be due to a selection bias, or to a redshift evolution of $\rm M_{UV}$ or \fesc\ \citep[e.g.][]{Mascia_24b, Cain_25b, Ferrara_25}. This would need to be studied further with mass-complete samples in the future. As described below, we will ignore any redshift evolution for the further analysis.

\subsection{UV magnitude dependence}

\begin{figure}
    \centering
    \includegraphics[width=\linewidth]{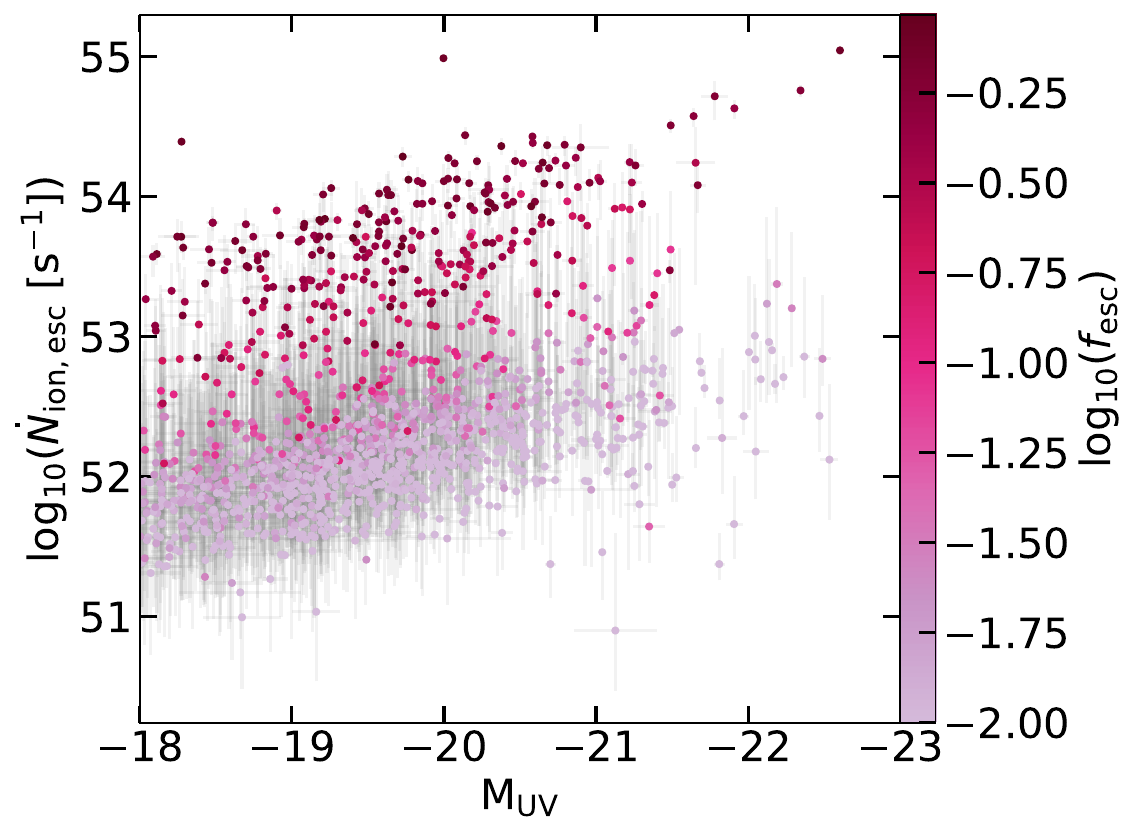}
    \caption{Distribution of $\dot{N}_{\rm ion, esc}$ as a function of UV magnitude color coded by $f_{\rm esc}$. At the same magnitude, sources with increased $f_{\rm esc}$ also show higher $\dot{N}_{\rm ion, esc}$, indicating that these are the sources that emit most ionizing photons into the IGM.}
    \label{fig:nion_vs_MUV}
\end{figure}

In Fig.~\ref{fig:nion_vs_MUV} we investigate the relation between $ \dot{N}_{\rm ion, esc}$ and UV magnitude. The ionizing photon production rate is proportional, to first order, to the rate of non-ionizing UV photon production. Hence, we expect $ \dot{N}_{\rm ion, int}$ to be proportional to the intrinsic UV luminosity. Indeed, it is immediately visible that $\dot{N}_{\rm ion, esc}$ increases towards brighter $\rm M_{UV}$. This allows us to investigate the impact of the escape mechanism on this relation. 
We color code the sample by \fesc, highlighting the same behavior already seen in Fig.~\ref{fig:ndot_vs_z}. All of the low \fesc\ sources are clustered around  ${\rm log}(\dot{N}_{\rm ion, esc}/\rm s^{-1})\sim 52$, while the $f_{\rm esc}\gtrsim10\%$ sources span the range $53 \lesssim {\rm log}(\dot{N}_{\rm ion, esc}/\rm s^{-1})\lesssim 55$, on average about 1.5 dex above the cloud of low \fesc\ sources. As mentioned above, however, the $\dot{N}_{\rm ion}$ values are artificially limited in our fitting because of the limit on $f_{\rm esc}$.

Overall, from Fig.~\ref{fig:nion_vs_MUV} it remains clear that galaxies with the highest escape fractions also emit the most ionizing photons into the IGM, in absolute number. Indeed, above $\rm log(\dot{\textit{N}}_{ion, esc}\, [s^{-1}])>53$ only 8\% of sources have an \fesc$<10\%$.
While faint galaxies with $f_{\rm esc}\sim 50\%$ may emit as many ionizing photons as bright galaxies with only $f_{\rm esc}\sim 10\%$, this result shows that $f_{\rm esc}$ is the key quantity to determine the sources of reionization, as it is the main factor controlling which sources leak a significant amount of ionizing photons. Indeed, even for very UV-bright sources that intrinsically produce a lot of ionizing photons, a very low escape fraction means that a negligible amount of them reach the IGM, even compared to fainter sources with an increased $f_{\rm esc}$. 


\section{Ionizing emissivity function}\label{LF}

\begin{figure}
    \centering
    \includegraphics[width = \linewidth]{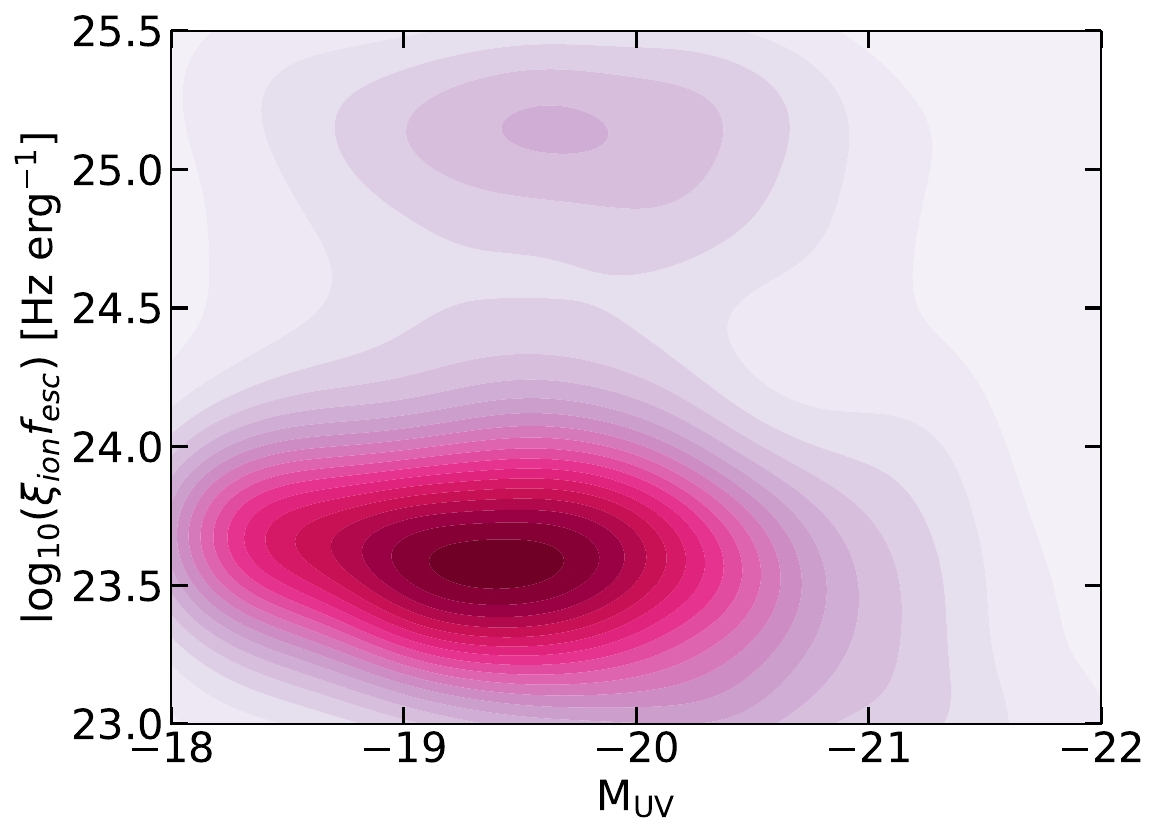}
    \caption{Kernel density estimation of ${\rm log}(\xi_{\rm ion}f_{\rm esc})$ as a function of $\rm M_{UV}$. The distribution shows negligible evolution as a function of $\rm M_{UV}$. Moreover, it also shows two clouds. The main one, with the majority of our data, includes all the weak leakers, corresponding to the pink points in Fig.~\ref{fig:nion_vs_MUV}. The secondary cloud, above the main one, includes the sources with high $f_{\rm esc}$, so the strong leakers.}
    \label{fig:KDE}
\end{figure}

We can remove the direct $\rm M_{UV}$ dependence by dividing $\dot{N}_{\rm ion, esc}$ by the intrinsic UV luminosity of each galaxy, leaving us with Fig.~\ref{fig:KDE}, where we show log($\xi_{\rm ion}f_{\rm esc}$) against $\rm M_{UV}$. Here we do no longer see the $\rm M_{UV}$ dependence, but we still see the difference between the strong and weak leakers.

In order to determine which sources contribute the most to reionization we must understand which sub-population of galaxies dominates the cosmic ionizing emissivity. To this goal, we construct a luminosity function of $\dot{N}_{\rm ion, esc}$, i.e., the ionizing emissivity function.
This function underpins the rest of our analysis: integrating it yields the cosmic emissivity $\dot{n}_{\rm ion}$, from which we derive the redshift evolution of the neutral fraction $x_{\rm HI}$ in Sect.~\ref{Neutral_fraction}. 

A similar approach has already been used by \cite{Mason_19} to estimate the contribution to $\dot{n}_{\rm ion}$ from quasars, albeit with a constant and unconstrained \fesc. We instead apply it to the general galaxy population and measure the escaping emissivity of each galaxy directly, so that \fesc\ need not be modeled separately.

Since our spectroscopic galaxy sample is assembled from various surveys, it is extremely complicated to derive a selection function. Hence, we derive the ionizing LF based on the UV LFs at these redshifts, convolved with our measurements of the $\rm log(\xi_{ion}\textit{f}_{esc})$ distribution. This is given by:
\begin{equation}
\begin{split}
\phi(\dot{N}_{\rm ion,esc}, z) &= \frac{dn(\dot{N}_{\rm ion,esc},z)}{d\dot{N}_{\rm ion,esc}} \\
&= \int_0^\infty dM_{\rm UV}\, \phi(M_{\rm UV},z) \, p(\dot{N}_{\rm ion,esc} \mid M_{\rm UV})
\end{split}
\end{equation}

Here, $p(\dot{N}_{\rm ion,esc} = L_{\rm UV}\xi_{\rm ion}f_{\rm esc} \mid M_{\rm UV})$ represents the distribution function of $\xi_{\rm ion}\textit{f}_{\rm esc}$ (shown in Fig. \ref{fig:KDE}) multiplied by the UV luminosity at a given $M_{\rm UV}$. 

We adopt this distribution at all redshifts: we cannot establish whether the mild redshift evolution of $\dot{N}_{\rm ion}$ seen in Fig.~\ref{fig:ndot_vs_z} is intrinsic or a selection effect. \cite{Simmonds_24b} find no redshift evolution in either $\dot{N}_{\rm ion}$ or $\xi_{\rm ion, 0}$ for a mass-complete sample, hence assuming no evolution for the general population is justified.

The second key ingredient for this calculation is thus an accurate measure of the UV LF across redshift. Of particular importance are the range $z=5$ to $z=15$, over which the process of reionization essentially completes \citep{Bosman_22}. 

For the UV LF, we adopt the measurements from the GLIMPSE survey \citep{Chemerynska_26, Atek_26} at $z>7$. As these are based on a deep lensing-cluster field, they constrain the faint end far better than blank-field surveys, which is essential here: assessing the contribution of the faintest galaxies requires integrating down to the faintest magnitudes. We adopt a faint-end cutoff of $\rm M_{UV}=-13$, the faintest bin reliably measured at these redshifts to date.

The GLIMPSE LFs of \cite{Chemerynska_26} are centered at $z = 9, 10, 11$ and $13$ and include photometric candidates up to $z\sim15$. They are fit with a double power law (DPL) rather than a Schechter function, since the latter underestimates the abundance of bright galaxies in the early Universe \citep{Adams_24, Donnan_24, Harikane_24, Chemerynska_26}. The lower-redshift GLIMPSE LF \citep{Atek_26} is centered at $z=7$, covering $6<z<9$, and is fit with a Schechter function. At $z=5$ and $6$ we use \cite{Bowler_20}, who adopt a DPL. For consistency across our full redshift range, we refit the $z=7$ LF points of \citep{Atek_26} with a DPL. The two parameterizations differ only in the brightest bins, so we expect this to have a negligible effect on our results.

The parameters of a UV LF are strongly correlated \citep[e.g.,][]{Bouwens_21, Atek_26}, so treating their uncertainties as independent would overestimate the true uncertainty on the LF. To propagate the uncertainties correctly into the ionizing emissivity function, we refit every LF with a DPL using the binned data points from \cite{Atek_26}, \cite{Chemerynska_26}, and references therein, which gives us access to the full parameter chains. Because we ultimately evaluate $\dot{n}_{\rm ion}$ at every integer redshift for a smooth interpolation, we also require the UV LF at every integer redshift. We obtain these by fitting a redshift-evolution function to each DPL parameter. The uncertainties are then calculated from the posterior distribution of the parameters. We report the DPL parameters and their uncertainties in Table~\ref{tab:DPL}.

We now have all the ingredients to construct the ionizing emissivity function, which we do as follows:
\begin{enumerate}
    \item we sample galaxies from the UV LF at a given redshift between $\rm M_{UV} = -13$ and $-23$, a range chosen to be comparable with other literature estimates \citep{Bouwens_15,Mason_18};
    \item we assign each $\rm M_{UV}$ a value of $\xi_{\rm ion}f_{\rm esc}$ drawn from the distribution in Fig.~\ref{fig:KDE}, which we apply at all magnitudes since our sample does not extend faintward of $\rm M_{UV}=-18$;
    \item we multiply the two to obtain $\dot{N}_{\rm ion, esc}$ for each source;
    \item we repeat steps (i)--(iii) $10^3$ times, drawing the UV LF parameters from the refitted chains to propagate the LF uncertainties.
\end{enumerate}

\begin{figure}
    \centering
    \includegraphics[width=\linewidth]{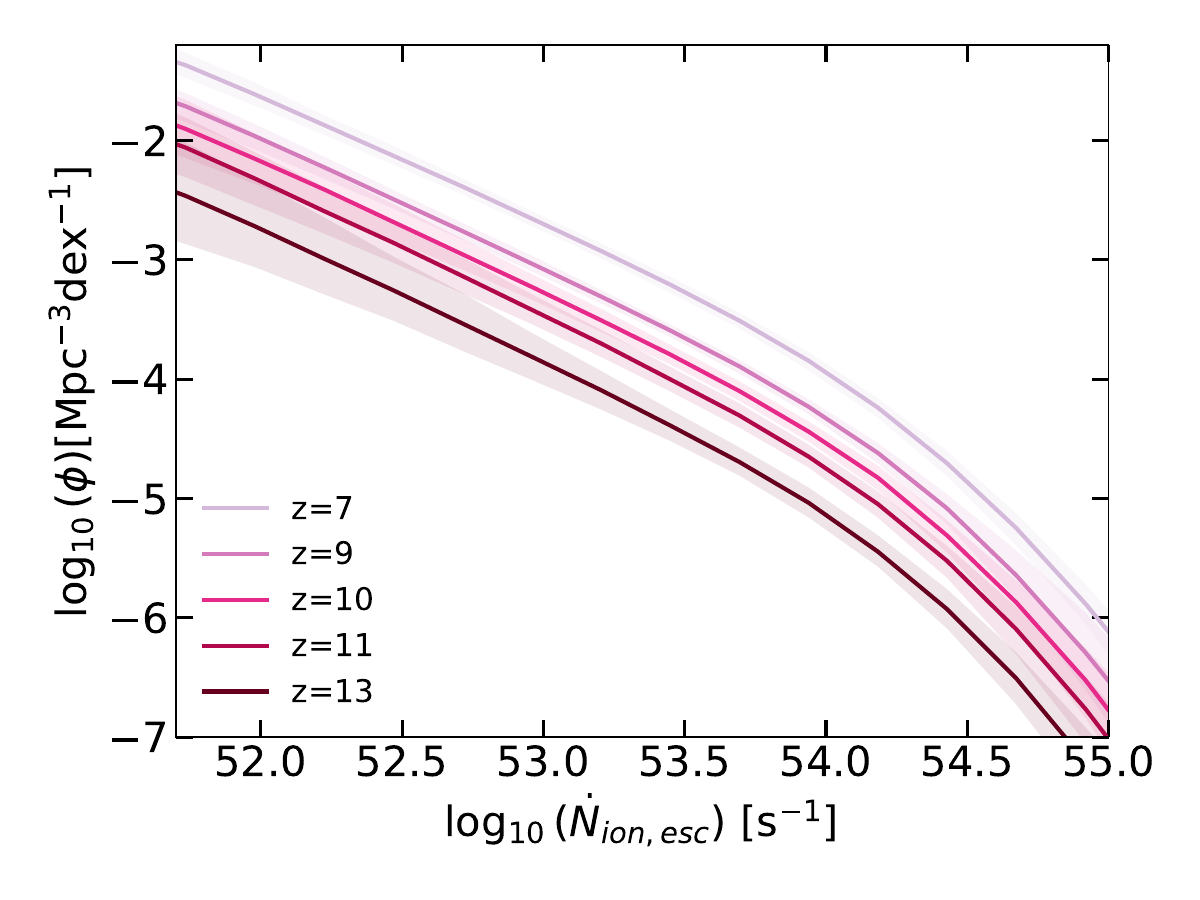}
    \caption{Ionizing emissivity function for five redshift ranges between $7<z<13$, derived from the UV luminosity function, in combination with our escaping ionizing emissivity on a galaxy-per-galaxy basis. The ionizing emissivity function decreases as redshift increases, in agreement with the evolution of the UV luminosity function.}
    \label{fig:ionizing_LF}
\end{figure}

The resulting functions are shown in Fig.~\ref{fig:ionizing_LF} for the redshifts with a GLIMPSE UV LF \citep{Chemerynska_26, Atek_26}; although we derive them at every integer redshift in $5<z<15$. The ionizing emissivity function increases towards lower redshift, mirroring the evolution of the UV LF and indicating that more ionizing photons reach the IGM per unit time at later epochs. This is expected, since $\dot{N}_{\rm ion, esc}$ scales with UV luminosity and we hold the distribution of $\rm log(\xi_{ion}\textit{f}_{esc})$ fixed with redshift by construction.

We now quantify the contribution of strong leakers (\fesc$>10\%$) to the ionizing emissivity function, to test whether the few sources that leak most of their ionizing photons indeed dominate it. This is not obvious a priori: although strong leakers release a large fraction of the photons they produce, they are in a minority, and the far more abundant weak leakers might collectively contribute more, despite their low \fesc\ of a few percent or less.

\begin{figure}
    \centering
    \includegraphics[width=\linewidth]{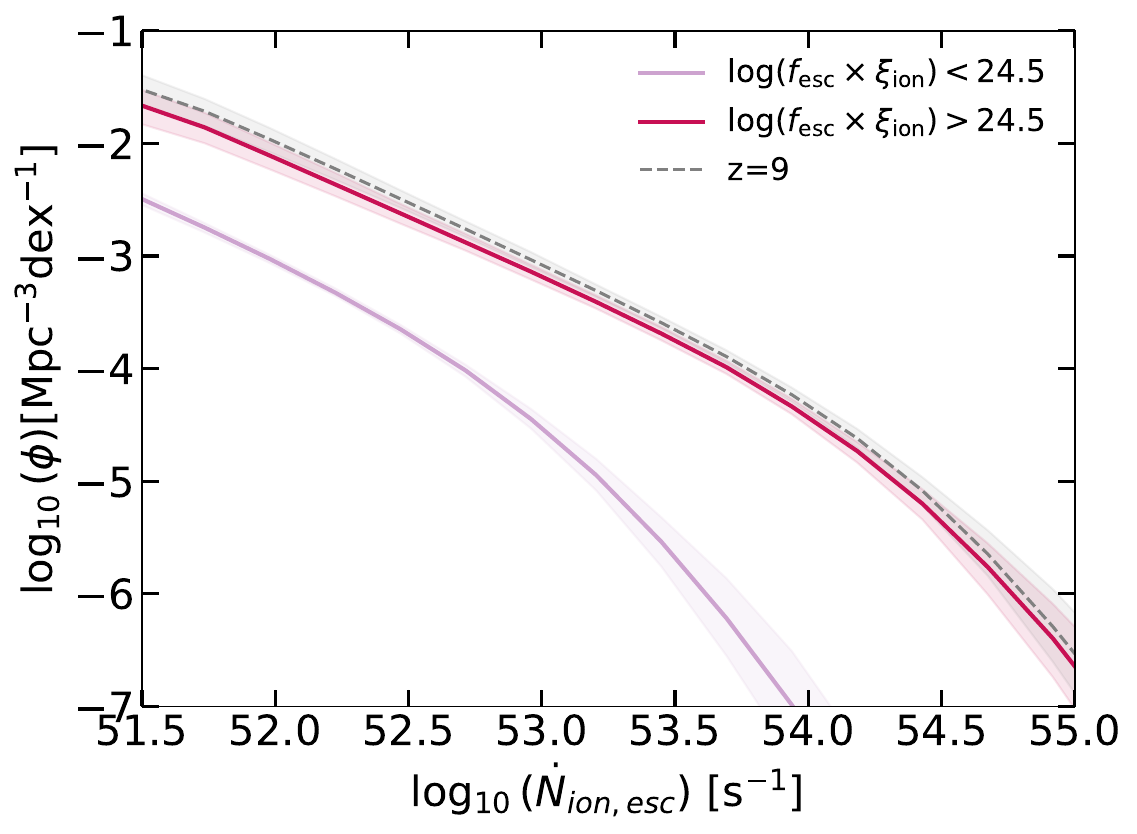}
    \caption{Ionizing emissivity function for redshift z=9, split by ${\rm log}(f_{\rm esc}\xi_{\rm ion})$ values. The gray dashed line is the total ionizing emissivity function, as seen also in Fig.~\ref{fig:ionizing_LF} while the two solid lines are the two partial emissivity functions, the red one being for ${\rm log}(f_{\rm esc}\xi_{\rm ion}/ \rm Hz \, erg^{-1}) > 24.5$ and the pink one being for ${\rm log}(f_{\rm esc}\xi_{\rm ion}/ \rm Hz \, erg^{-1}) < 24.5$} 
    \label{fig:ionizing_LF_split}
\end{figure}

Fig.~\ref{fig:ionizing_LF_split} shows the same ionizing output luminosity function as in Fig.~\ref{fig:ionizing_LF} for redshift z = 9, but split at ${\rm log}(f_{\rm esc}\xi_{\rm ion}/\rm Hz \, erg^{-1}) = 24.5$, which corresponds to \fesc$\sim10\%$ and thus separates strong leakers from weak leakers. The total emissivity is dominated by the high-\fesc\ sources above this threshold, while the contribution of the rest is negligible, especially at the highest values of $\rm log(\dot{\textit{N}}_{ion, esc})$. We therefore confirm that, despite being far more numerous, sources with \fesc$<10\%$ contribute negligibly to the rate of ionizing photons reaching the IGM. These strong leakers make up only $\sim20\%$ of our sample.

\section{Neutral gas fraction}\label{Neutral_fraction}

\begin{figure}
    \centering
    \includegraphics[width=\linewidth]{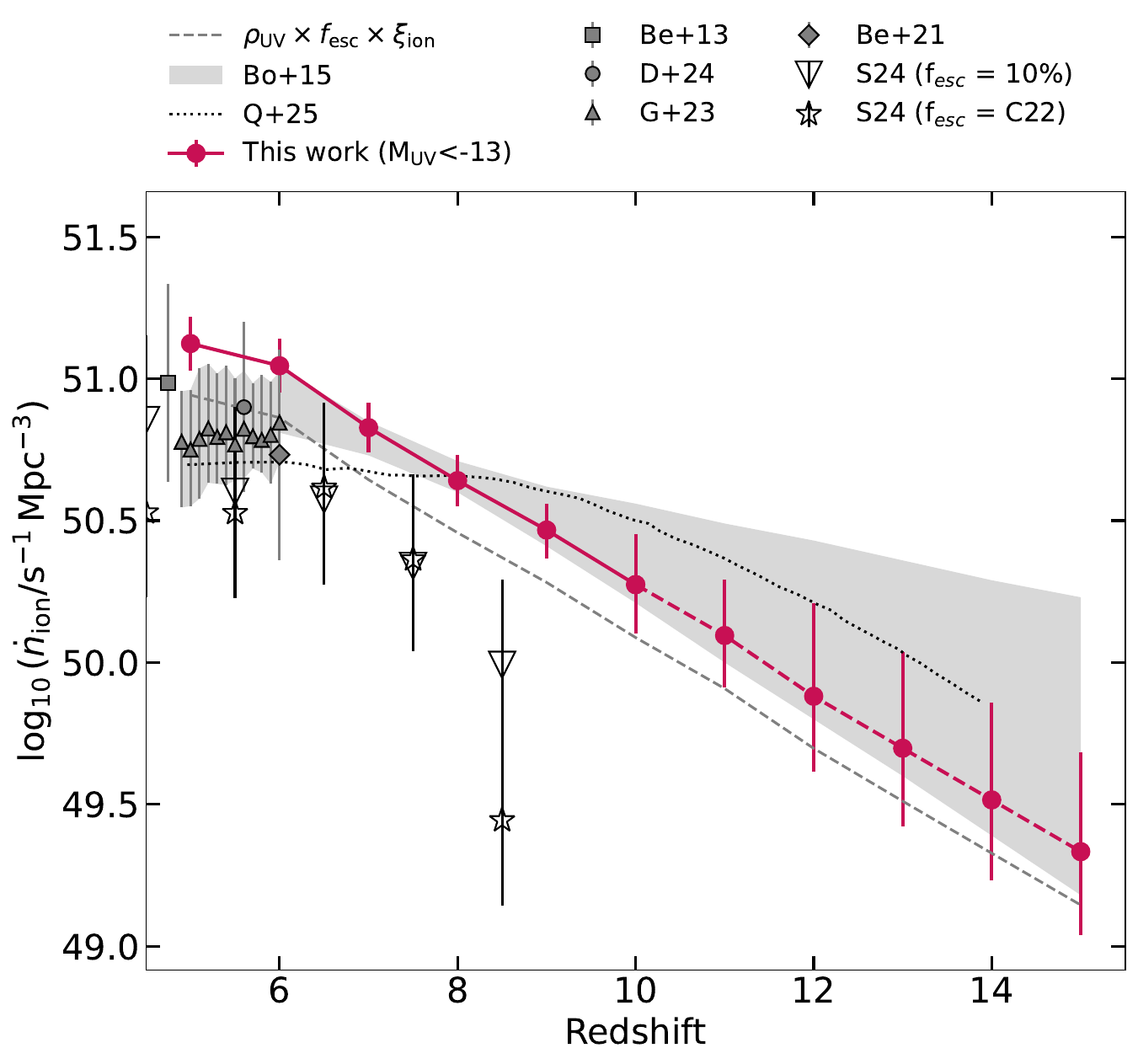}
    \caption{Cosmic rate of ionizing photons emission per unit time and unit volume. We compare our results from other observational constraints using quasars Lyman-$\alpha$ forest \citep{Becker_13, Becker_21, Gaikwad_23, Davies_24, Qin_24} and UV luminosity functions \citep{Bouwens_15, Simmonds_24b}, assuming a clumping factor $C_{\rm HII} = 3$. The dashed grey line is derived by treating $\rho_{\rm UV}$, $f_{\rm esc}$, $\xi_{\rm ion}$ independently. This assumes the luminosity density from the GLIMPSE UV LFs and the means of the parameters we find in this study ($f_{\rm esc} = 10\%$ and $\rm log(\xi_{\rm ion}/ Hz
    \, erg^{-1}) = 25.45$)}
    \label{fig:nion_vs_z}
\end{figure}

The most commonly used probes of the global timeline of reionization are the volume-filling fraction of ionized gas, $Q_{\rm HII}(z)$, and its counterpart, the fraction of neutral hydrogen ($x_{\rm HI} = 1- Q_{\rm HII}$). To determine the fraction of neutral hydrogen and its evolution during the EoR we need to know the contribution of ionizing photons to the IGM by the entire population of galaxies, $\dot{n}_{\rm ion}$. This quantifies the rate of ionizing photon escape from galaxies per unit time and unit volume and is usually defined as Eq.~\ref{eq:nion}.

Again we do not attempt to disentangle the contributions of $f_{\rm esc}$ and $\xi_{\rm ion}$ to avoid adding unnecessary sources of uncertainty. We simply calculate $\dot{n}_{\rm ion}$ by integrating the entire ionizing emissivity function as
\begin{equation}
    \dot{n}_{\rm ion} = \int \dot{ N}_{\rm ion}\phi(\dot{N}_{\rm ion})d\dot{N}_{\rm ion}
\end{equation} 
which by construction of our $\rm M_{\rm UV}$ sampling, is analogous to integrating down to $\rm M_{UV} = -13$. This limit is chosen as it has been already used in the literature \citep{Bouwens_15} to include the faintest sources while avoiding a possible turnover in the UV LF. In Fig. \ref{fig:nion_vs_z} we show the evolution of our inferred cosmic rate of ionizing photon emission as a function of redshift. At low redshift we find ourselves largely in agreement with the uncertainties of observational constraints on $\dot{n}_{\rm ion}$ from quasar Lyman-$\alpha$ forests \citep{Becker_13, Becker_21, Davies_24}, although at $z=6$ we are in slight tension ($\sim 1\sigma$) with the observational constraints of \cite{Gaikwad_23}, possibly due to their assumptions on mean free path and photo-ionization rate. At intermediate redshifts we are also in agreement with the results of \cite{Simmonds_24b}. These are obtained from the \cite{Bouwens_21} UV LF and two different models for \fesc, one assuming a constant $10\%$ and one assuming the relation from \cite{Chisholm_22}. Above $z>6$ we are entirely in agreement with the estimates from \cite{Bouwens_15}, who assume a clumping factor $C_{\rm HII} = 3$, as we do. At $z>8$ we find a broad agreement with the models based on XQR-30 data from \cite{Qin_24}. In the same figure we also show the evolution of $\dot{n}_{\rm ion}$ we would get by treating the three factors independently (i.e. using Eq.~\ref{eq:nion}). For this calculation we use the mean parameters we find from our SED fitting ($f_{\rm esc} = 10\%$ and $\rm log(\xi_{\rm ion}/Hz \, erg^{-1}) = 25.45$). We find these results to be more in agreement with those from \cite{Simmonds_24b} and the Lyman-$\alpha$ forest \citep{Becker_13, Becker_21, Davies_24}, while not being consistent with results from \cite{Bouwens_15} and \cite{Qin_24}.
This shows that it is necessary to account for the correlation of the parameters in this calculations as it significantly affects the inferred cosmic rate of ionizing photons emission. 

From $\dot{n}_{\rm ion}$ it is relatively straightforward to compute the volume-filling fraction of ionized gas by solving the differential equation 
\begin{equation}\label{eq:reionization}
   { \rm \dot{Q}_{HII}} = \frac{\dot{n}_{\rm ion}}{\langle n_{\rm H}\rangle} - \frac{\rm Q_{HII}}{t_{\rm rec}}
\end{equation}
where $\langle {n}_{\rm H} \rangle$ is the average hydrogen number density and $t_{\rm rec}$ is the recombination time. $t_{\rm rec}$ is defined as 
\begin{equation}
    t_{\rm rec} = [C_{\rm HII}\alpha_{\rm B}(T)(1+Y_{p}/4X_{p})\langle n_{\rm H}\rangle (1+z)^3]^{-1}.
\end{equation}
For this calculation we assume a recombination coefficient for a temperature of $\rm T = 10^4 \, K$ of $\alpha_{\rm B}(T) = 2.6\times10^{-13} \, \rm cm^3 s^{-1}$ from \cite{Osterbrock_06}, primordial abundances of hydrogen and helium such that $\rm 1+Y_{p}/4X_{p} = 1.08$ and a clumping factor $\rm C_{HII}=3$, motivated by results from simulation \citep{Finaltor_12, Shull_12}. It should be noted that other studies have shown that the clumping factor may be higher than $\rm C_{HII}=3$. \cite{Austin_25} have shown that, to reionize the Universe by $z=6$, the clumping factor may be 6.2 or even as high as 8.8, depending on the model used for $f_{\rm esc}$. It has also been shown from models and simulations that $\rm C_{HII}$ is likely to evolve, increasing from $\sim 1$ to $\sim 8$ as reionization proceeds \citep{Pawlik_15, Qin_24}. An increase of clumping factor would affect our $x_{\rm HI}$ calculations by postponing the end of reionization by $\Delta z = 0.4$ for $\rm C_{HII} = 6$.

\begin{figure*}
    \centering
    \includegraphics[width=\linewidth]{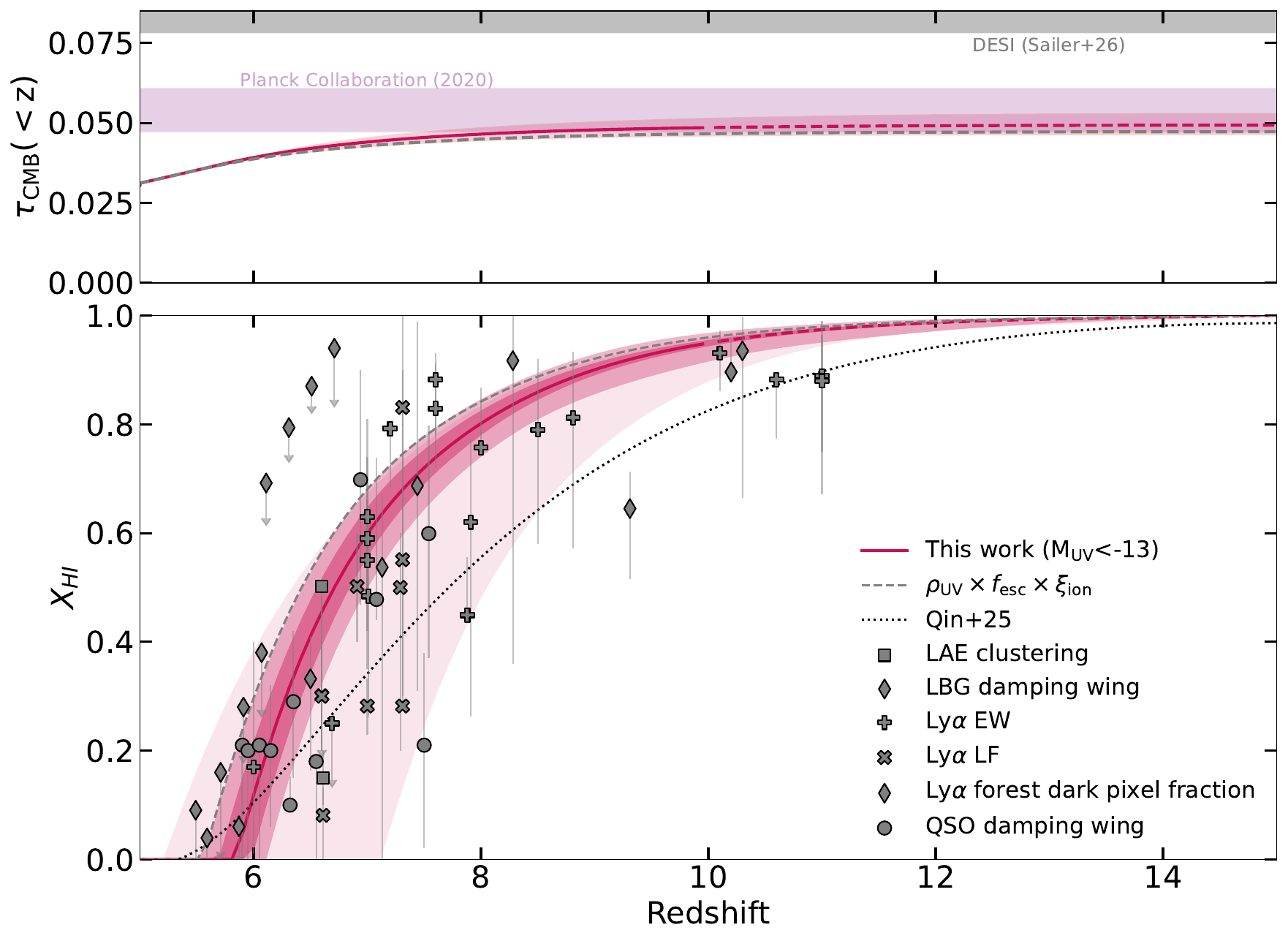}
    \caption{\textbf{Top panel:} Evolution of the optical depth of electron scattering as a function of redshift. Our results agree with constraints from \cite{Planck_18}, in the pink shaded area, but are much lower than recent values hypothesized by \cite{DESI_26}, based on DESI data, to solve the neutrino mass tension, shown in grey. \textbf{Bottom panel:} Neutral hydrogen fraction as a function of redshift, compared to literature points. Our results, down to an integration limit of $\rm M_{UV} = -13$ are shown with the red line, where the dashed section represents the redshift range in which we have extrapolated our results. The lightest shaded area represents the uncertainty when using the UV LF from \cite{Donnan_24}. The middle shaded area represents the uncertainty using the GLIMPSE LFs \citep{Chemerynska_26, Atek_26}. The smallest and darkest shaded area represents the uncertainty if we assume no uncertainty on the UV LF, indicating that the this is in fact the largest source of uncertainty in this calculation. The gray dashed line shows the evolution estimated with just average values, i.e. $\rho_{\rm UV}$ from the UV LF, $f_{\rm esc} = 10\%$ and $\xi_{\rm ion} = 25.45$, the means of the parameters as found from the SED fitting. The gray points show literature estimates of $x_{\rm HI}$ from various probes (see main text for references). 
    The dotted black line shows the neutral hydrogen fraction derived in \citep{Qin_24}.}
    \label{fig:xHI}
\end{figure*}

Solving Eq.~\ref{eq:reionization} we find the evolution of $Q_{\rm HII}$ and therefore of $x_{\rm HI}$ with redshift. Our results are presented in Fig.~\ref{fig:xHI}. Here we compare them with other literature values for a variety of probes of the ionization state of the IGM such as the Lyman-$\alpha$ emitters clustering \citep{Sobacchi_15, Ouchi_18}, Lyman-break galaxies damping wing \citep{Hsiao_24, Umeda_24, Mason_26}, the Lyman-$\alpha$ equivalent width evolution \citep{Mason_18,Mason_19b, Hoag_19, Whitler_20, Bolan_22, Bruton_23, Morishita_23, Nakane_24, Tang_24, Kaguera_25}, Lyman-$\alpha$ luminosity functions \citep{Ouchi_10, Konno_14, Konno_18, Zhengh_17, Inoue_18, Goto_21, Morales_21, Ning_22}, Lyman-$\alpha$ forest dark pixel fraction \citep{Jin_23}, quasar damping wings \citep{Schroeder_13, Davies_18, Wang_20, Greig_19, Greig_24} and the Lyman-$\alpha$ forest \citep{Qin_24}. Overall we find our results to be broadly consistent with these estimates. We find an end to reionization at $z\sim5.8$, also in agreement with recent estimates from quasar spectra such as \cite{Bosman_22}, who find an end to reionization at $z\sim 5.3$ but $x_{\rm HI} \sim 6\times10^{-5}$ at $z\sim 5.8$. 

Surprisingly, we find that the largest source of uncertainty in the calculation of the neutral hydrogen fraction in this framework is not the uncertainty on $f_{\rm esc}$ and $\xi_{\rm ion}$, but that on the UV luminosity density; in particular, the uncertainty on the faint end slope of the LF. In Fig.~\ref{fig:xHI} we show three shaded regions, each representing a different uncertainty. The smallest shaded area represents the uncertainty which does not consider any uncertainty in the UV LF, using only the best fit parameters. The second shaded area shows the uncertainty in the estimation of $x_{\rm HI}$ when we also consider the uncertainty in the GLIMPSE LF. Lastly, we use the same method detailed in the previous section with the LF parameters from \cite{Donnan_24}. These LFs are much less constrained at the faint end. This is due to the lack of very faint sources, as their data does not include very deep and lensed fields such as the GLIMPSE field. In this case, the uncertainty on $x_{\rm HI}$ significantly increases, showing that constraining the faint end of the LF at high redshift is necessary to determine the timeline and evolution of reionization. The values of $x_{\rm HI}$ with the associated uncertainties for both GLIMPSE and \cite{Donnan_24} are tabulated for all integer redshifts in Tab.~\ref{tab:xHI}. 

We also compare the redshift evolution of the neutral fraction to the standard estimate that only uses average values rather than the actual distribution functions, i.e., $\dot{n}_{\rm ion} = \rho_{\rm UV}f_{\rm esc}\xi_{\rm ion}$. For the UV luminosity density we again integrate the GLIMPSE LF, while for other quantities we use the mean values for our samples of $f_{\rm esc} = 0.1$ and $\rm log(\xi_{\rm ion}/ \rm Hz \, erg^{-1}) = 25.45$.
As already discussed, treating these parameters independently significantly affects the reionization calculations. This is also seen here, where we find a delayed progress of reionization ending at $z\sim5.4$. As all the quantities in the calculation are derived from the same information in the spectrum, the line fluxes and the UV slope, they must not be treated independently, to avoid underestimating the ionizing $\dot{n}_{\rm ion}$ and to accurately describe reionization. 

It is also possible to compare this reionization history with cosmological constraints by measuring the optical depth of electron scattering as measured from the CMB $\tau_{\rm CMB}$. It is defined as \citep{Robertson_22}:
\begin{equation}
    \tau(z) = c\langle n_{\rm H}\rangle \sigma_{\rm T} \int_0^{z} f_{e} Q_{\rm HII}(z')H^{-1}(z')(1+z')^2dz'
\end{equation}
where $c$ is the speed of light, $\langle n_{\rm H} \rangle$ is the comoving average hydrogen number density, $\sigma_{\rm T}$ is the Thomson cross-section, $H(z)$ is the redshift dependent Hubble parameter and $f_{e}$ the number of free electrons per hydrogen atom, which depends on the ionization state of hydrogen and helium. We find a value of $\tau_{\rm CMB}$ consistent with results from \cite{Planck_18} but much lower than a recent value suggested by \citep{DESI_26}, based on DESI data, which would solve the neutrino mass tension. This value of $\tau$ is itself in a $\sim 2\sigma$ tension with other estimates \citep{Cain_25}.

\subsection{Contributions to the neutral gas fraction}

\begin{figure}
    \centering
    \includegraphics[width=\linewidth]{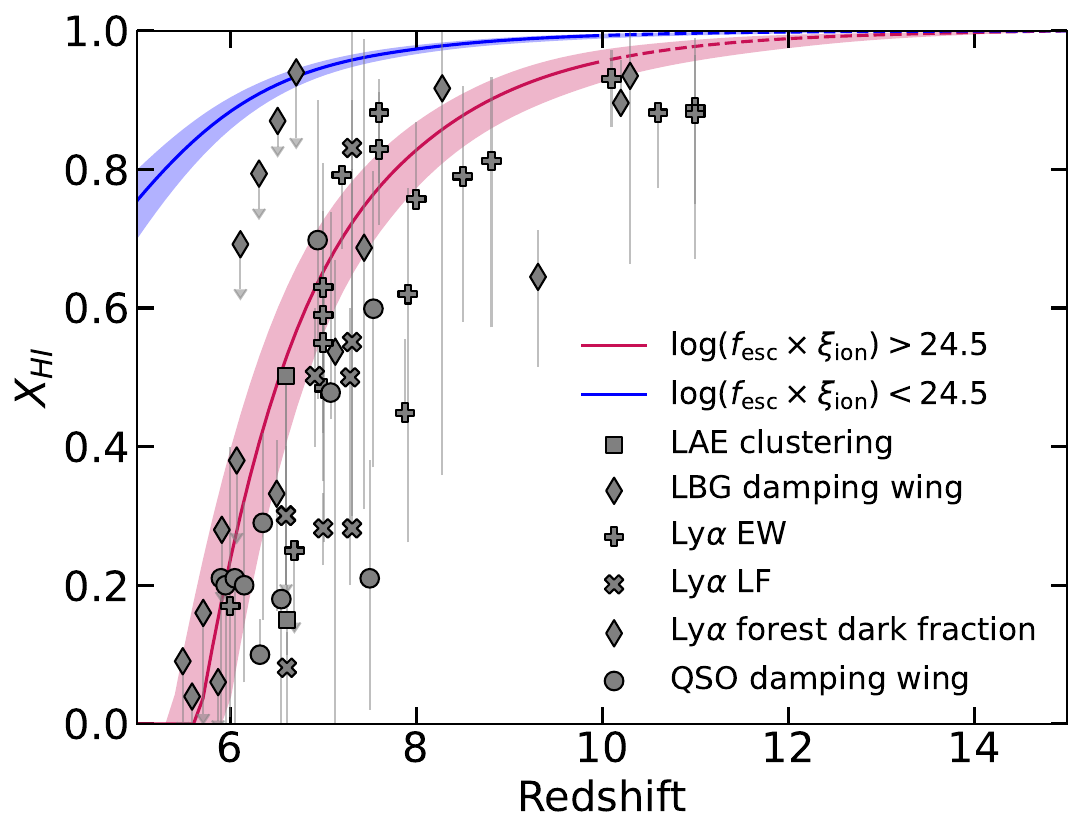}
    \caption{Neutral hydrogen fraction as a function of redshift split at ${\rm log}(f_{\rm esc}\xi_{\rm ion}/\rm Hz \, erg^{-1}) = 24.5$. The red line only includes the sources with ${\rm log}(f_{\rm esc}\xi_{\rm ion})$ above this value while the blue line only includes sources below this value. This shows that strong leakers ($f_{\rm esc} \gtrsim 0.1$) alone do the majority of the reionization process, while the contribution of the remaining sources is not significant.}
    \label{fig:xHI_split}
\end{figure}

One of the biggest open questions in the study of the EoR is understanding which kind of sources contribute most to the reionization process. We try to answer this question by quantifying the contribution of strong leakers to the evolution of the neutral hydrogen fraction over the history of the Universe. This will tell us whether it is the few highly leaking sources or the many weakly leaking sources that drive the bulk of the reionization process. 
Similarly to what was already done in Fig.~\ref{fig:ionizing_LF_split} we split our sampling between sources with log($\xi_{\rm ion}f_{\rm esc}/ \rm Hz \, erg^{-1})< 24.5 $ and those above that value and calculate what the evolution of the neutral hydrogen fraction for each sample (see Fig \ref{fig:xHI_split}). We find that reionization is almost entirely driven by the few sources with a high escape fraction. These sources alone can ionize the universe in agreement with the observational constraints. The majority of sources, those with $f_{\rm esc} \lesssim 0.1$, do not have a significant contribution to the ionization of neutral gas in the universe. 
This shows that it is in fact the few sources with very high escape fraction, $\sim 20\%$ of the total sample, that perform the bulk of reionization, producing $\sim 87\%$ of all ionizing photons in the EoR, instead of the majority of sources with low escape fraction. This is broadly in agreement with recent results from the PRISMS survey, who also find that a minority of sources with very high \fesc\ should be responsible for the majority of the reionization process \citep{Marques_chaves_26}. It should be noted that high \fesc\ is likely a temporal event. Simulations in fact show that \fesc\ can drastically change in galaxies as a function of time, usually following star formation events \citep{Trebitsch_17}. It would therefore not be the same 20\% of sources driving all of the reionization process but nonetheless, at any given time, we expect that only a minority of sources strongly contribute to this process.

\subsection{Do faint or bright galaxies drive reionization?}

\begin{figure*}
    \centering
    \includegraphics[width=\linewidth]{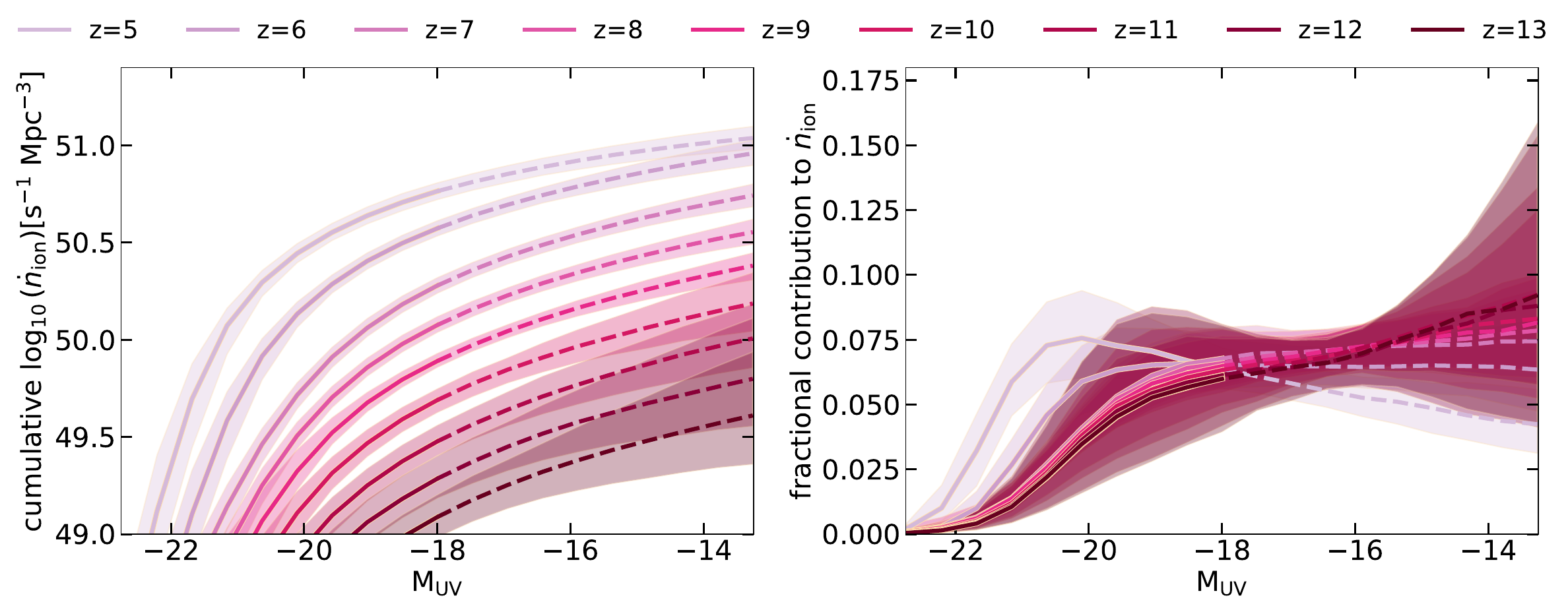}
    \caption{\textbf{Left panel}: cumulative distribution of $\dot{n}_{\rm ion}$ as a function of $\rm M_{UV}$ for nine integer redshift in the range $5<\rm z < 13$. For all the lines, the dashed section represents the magnitude range in which we have extrapolated our results, as our initial sample only includes sources with $\rm M_{UV}<-18$. Overall, the amount of ionizing photons emitted per unit time and unit volume increases at lower redshift. Also, the contribution of fainter sources at lower redshift becomes less important compared to bright sources. \textbf{Right panel}: fractional contribution to the total $\dot{n}_{\rm ion}$ at each magnitude for the same redshifts, in bins of $\Delta \rm M_{UV} = 0.5$. The dashed sections once again represent the range in which our results are extrapolated. While at high redshift, faint sources ( $\rm M_{UV}\gtrsim-18$) contribute the most to the total $\dot{n}_{\rm ion}$, at low redshift bright sources contribute the most.}
    \label{fig:cumulative}
\end{figure*}

It is still unclear at what magnitude range galaxies contribute the most to the reionization, whether it is faint \citep{Atek_24} or bright galaxies \citep{Naidu_20}. Recent work targeting the faintest galaxies finds that those do not have the highest escape fractions and that the largest contribution of ionizing photons comes from galaxies with $\rm M_{UV}\sim -16$ \citep{Jecmen_26}.
In Fig.~\ref{fig:cumulative} we investigate which sources contribute the most to the reionization process. First, in the left panel, we show the cumulative distribution of $\dot{n}_{\rm ion}$ as a function of $\rm M_{UV}$ for nine integer redshifts between $5< z < 13$.
Overall we see that the amount of ionizing photons emitted into the IGM increases as redshift decreases, as we already saw in Fig.~\ref{fig:ionizing_LF}. We also see that in general very bright galaxies do not contribute strongly to the total amount of ionizing photons in a unit volume. Moreover, we find that the cumulative distribution flattens at different magnitudes for the different redshift. We note that as we have assumed no dependence of $\xi_{\rm ion}f_{\rm esc}$ on $\rm M_{UV}$ these results would be the same for non ionizing UV radiation. This indicates that at higher redshifts faint galaxies have a larger contribution than at the end of reionization, where we find the distribution already flattening at $\rm M_{UV}\sim -18$.
In Fig.~\ref{fig:cumulative}, we also show the normalized fractional contribution of sources in bins of 0.5 $\rm M_{UV}$ to the total $\dot{n}_{\rm ion}$. Here we can again see that the fractional contribution varies with redshift, with bright sources being more important at lower ($z \leq 6$) redshifts. Indeed, for the case of $z=5$ we find that the largest contribution to the total $\dot{n}_{\rm ion}$ comes from galaxies with $\rm M_{UV}\sim -20$. This is also true for $z=6$ although the contribution of fainter sources is more significant than at the lowest redshift. Above $z>6$ we find that bright galaxies do not strongly contribute to the total $\dot{n}_{\rm ion}$ while fainter sources, with $\rm M_{UV}>-19$ equally contribute. This trend is consistent with the constant $f_{\rm esc} = 14\%$ from \cite{Jecmen_26}. While their fiducial distribution shows a peak at $\rm M_{UV} = -16$, which ours doesn't, they also find that sources with $\rm M_{UV}$ between -18 and -14 drive reionization, which is consistent with our results, within the limits of extrapolation. It is also in agreement with recent results from the THESAN-ZOOM simulation, which show that brighter galaxies contribute more at lower redshifts \citep{Summerfield_26}. 

Overall, this tells us that faint galaxies contribute more at high redshifts while at lower redshifts their contribution decreases in favor of brighter galaxies. It is important to note that we do not assume any magnitude evolution for $\rm log(\xi_{ion}\textit{f}_{esc})$, effectively assuming that faint galaxies are as efficient in emitting ionizing photons as bright galaxies. The reason why they dominate the contribution of ionizing photons to the IGM is that, although they individually produce less ionizing photons, they are much more numerous, especially at high redshift, where very bright sources are increasingly rarer. This trend is therefore highly dependent on the faint end slope of the luminosity function. Indeed, if we compare our results to those based on [O\textsc{III}]+H$\beta$ luminosity function we find that they are inconsistent. Both \cite{Korber_26} and \cite{Meyer_24} find that, when deriving $\dot{n}_{\rm ion}$ from the [O\textsc{III}]+H$\beta$ luminosity function, bright galaxies contribute more at higher redshifts. This discrepancy is due to the faint end of the [O\textsc{III}]+H$\beta$ and UV luminosity functions diverging at high redshift, with the UV LF having a steeper slope, increasing the contribution of faint sources in our model. This discrepancy reinforces the need for better constraints on the faint end slope of UV LFs at high redshift. 

\section{Caveats}\label{Discussion}

Here we would like to discuss the main assumptions and caveats of this work. 

First of all we would like to once again note, as already discussed in \cite{Giovinazzo_25}, the degeneracy between star formation history and escape fraction, as this would affect our estimates of $\dot{N}_{\rm ion, esc}$. However, we had already found that this uncertainty is less important for very high \fesc sources, which often have spectra that cannot be reproduced with \fesc= 0. Therefore, as we have found that the sources with $f_{\rm esc}>10\%$ are the main drivers of reionization, we do not think that this degeneracy will have a strong impact on our results. Moreover, there is a category of sources, remnant leakers \citep{Katz_23}, for which this degeneracy is not important, as they are quenched sources with nonetheless high \fesc. These sources would have still young stellar populations but little ISM, making them efficient leakers. 

We would also like to note that since our method to calculate $\dot{N}_{\rm ion, esc}$ relies on SED fitting, it includes assumptions on the stellar populations, initial mass functions (IMFs) and dust attenuation laws. We assume a BPASS v2.2.1 stellar population \citep{Stanway_Eldridge_18} with the default broken power law IMF, so if these assumptions were to not be accurate it could affect our results. In this context, our quantitative conclusions should be interpreted within our modeling framework, although we do not believe that our qualitative conclusions on the sources of reionization would be affected. 

Possibly the biggest caveat of the analysis presented in this work is the assumption of a lack of magnitude evolution for $\rm log(\xi_{ion}\textit{f}_{esc})$. This assumption allows us to expand the analysis to magnitudes not covered by our initial sample and investigate the impact of the faintest galaxies to reionization. However, just because we find no evidence of a trend between $\rm log(\xi_{ion}\textit{f}_{esc})$ and $\rm M_{UV}$ in our sample it does not necessarily mean that this will be true at all magnitudes. Overcoming this would require an analysis of the ionizing output of sources covering the fainter end of the magnitude distribution that we are lacking here. Moreover, due to our choice to use only spectra with a confident redshift estimate, we might be biasing our sample to only strong line emitters, which would also bias our $\rm log(\xi_{ion}\textit{f}_{esc})$ distribution.

Lastly, we would like to mention the issue of cosmic variance. As we have seen that the estimate of the neutral hydrogen fraction is very sensitive to the UV luminosity function uncertainty, especially in the faint end slope, we have chosen to use the luminosity function from GLIMPSE \citep{Chemerynska_26, Atek_26}. However, GLIMPSE is only one NIRCam pointing of a lensed cluster, which significantly reduces the effective volume. It could therefore be highly affected by cosmic variance, although this effect is less problematic for very faint sources. Alleviating this issue would require many more extremely deep field observations with \textit{JWST} or the upcoming \textit{Extremely Large Telescope} (ELT).

\section{Conclusion}\label{Conclusion}

In this work we leverage JWST/NIRSpec spectra from a variety of public observations in the redshift range $5<z<10$, retrieved from the DJA, to compile a sample of 1428 sources. Using the SED fits already performed in \cite{Giovinazzo_25} we investigate the ionizing photon production of galaxies in the Epoch of Reionization. We estimate the ionizing photon output of individual galaxies, accounting for their intrinsic ionizing photon production and their escape fraction. Starting from the GLIMPSE UV luminosity functions \citep{Chemerynska_26, Atek_26}, as a way to account for the incompleteness of our sample, we also compute the ionizing emissivity function. This quantifies the amount of sources emitting a given rate of ionizing photons in the IGM per unit time per unit volume. We also determine the importance that high $f_{\rm esc}$ sources ($f_{\rm esc}>0.1$) have for the reionization process and whether, although not very numerous, they drive it. Our main findings are summarized below.

\begin{itemize}
    \item From our SED fits we calculate the amount of ionizing photon produced by each galaxy, as well as the amount of ionizing photons emitted into the IGM, when the escape fraction is taken into account. We find that the intrinsic production shows a slight evolution towards higher values at higher redshifts. This is consistent with the results of \cite{Simmonds_24b}, who also use a spectroscopic only sample, possibly indicating that this trend is due to the incompleteness and selection function of the sample. We find a similar trend for the escaped rate of ionizing photons, which also increases as a function of redshift, although at a different rate for $f_{\rm esc}<0.1$ and $f_{\rm esc}>0.1$. Indeed we find a stronger evolution, more consistent with the intrinsic, for strong leakers, while weak leakers show an almost flat trend. 
    \item We find that in general the sources with higher $f_{\rm esc}$ are in fact those that emit the most ionizing photons into the IGM, which is an expected although not obvious result. At a constant UV magnitude we find a cloud of low $f_{\rm esc}$ sources about $\sim 1.5$ dex below the leakers. This translates into the presence of two clouds, one of strong leakers and one of weak leakers also in log$_{10}(\xi_{\rm ion} f_{\rm esc}$).
    \item The ionizing emissivity function decreases with increasing redshift, following the behavior of the UV luminosity function. We study the behavior of this function when splitting the sample at log$_{10}(\xi_{\rm ion}f_{\rm esc}$) = 24.5, roughly corresponding to $f_{\rm esc} \sim 0.1$ split and the mean of the sample. In this case we find that the main contribution to the rate of ionizing photons emitted into the IGM per unit volume comes almost entirely from sources above these values. Only at low $\dot{N}_{\rm ion}$ values low $f_{\rm esc}$ sources start contributing. 
    \item Based on the ionizing emissivity function we estimate the evolution of the fraction of neutral hydrogen with redshift. We find its evolution to be consistent with literature estimates and that the uncertainty on the UV luminosity function is the lead cause of uncertainties in our estimates, especially the uncertainty in the faint end slope. We compare this evolution to that found when assuming $f_{\rm esc} = 0.1$ and $\rm log(\xi_{\rm ion} / \rm \, Hz \, erg^{-1}) = 25.45$ as motivated from our results. We find them to be in broad agreement with the latter being faster. We also estimate $\tau_{\rm CMB}$ and find it in agreement with the results from \cite{Planck_18}. 
    \item We determine the contribution to the ionized hydrogen fraction by strong leakers by performing again the same analysis over two samples split at log($\xi_{\rm ion}f_{\rm esc}/  \rm \, Hz \, erg^{-1}$) = 24.5. We find that the strong leakers ($f_{\rm esc}>0.1$) alone are almost able to match the reionization history that we find for all the sources. This shows that the few strongly leaking sources, $\sim$20\% of the total sources in the universe, are able to entirely drive the reionization process, producing $\sim 87\%$ of all ionizing photons that reach the IGM. This indicates that it is necessary to identify highly leaking galaxies to find the drivers of reionization.
    \item Lastly, we study the contribution to the total $\dot{n}_{\rm ion}$ from sources of different magnitudes. We find that at high redshift sources with $\rm M_{UV} > -18$ all seem to contribute equally, while at low redshift bright sources ($\rm M_{UV}\sim -20$) contribute most.
\end{itemize}

With this work we have presented a framework to calculate the ionizing photon emission of individual galaxies, as well as the rate of ionizing photons being emitted in a unit volume. Our results are consistent with observational constraints and show that 20\% of galaxies with high $f_{\rm esc}$ are almost entirely responsible for reionization, rather than the majority of low ionizing output galaxies.
The relative contributions of faint and bright galaxies vary with redshift, with
bright sources dominating at the end of reionization and faint sources at its
onset. Pinning down the drivers of reionization will ultimately require pushing
$f_{\rm esc}$ measurements to the faint, abundant galaxies that our spectroscopic
sample cannot yet reach, a regime that deep lensed-field spectroscopy with
JWST, and eventually the ELT, is poised to study in the future.

\begin{acknowledgements}
    
This work is based on observations made with the NASA/ESA/CSA James Webb Space Telescope. The raw data were obtained from the Mikulski Archive for Space Telescopes at the Space Telescope Science Institute, which is operated by the Association of Universities for Research in Astronomy, Inc., under NASA contract NAS 5-03127 for \textit{JWST}. 
Some of the data products presented herein were retrieved from the Dawn JWST Archive (DJA). DJA is an initiative of the Cosmic Dawn Center, which is funded by the Danish National Research Foundation under grant No. 140 (DNRF140).
      
This work has received funding from the Swiss State Secretariat for Education, Research and Innovation (SERI) under contract number MB22.00072, as well as from the Swiss National Science Foundation (SNSF) through project grants 200020\_207349 and 2000-1-243073.
\end{acknowledgements}

\bibliographystyle{aa}
\bibliography{bib}
\appendix
\onecolumn

\section{UV LF parameters}
In Table~\ref{tab:DPL} we compile the parameters and related uncertainties of the double power law parameters used in the analysis.

\begin{table}[h!]
    \centering
    \caption{DPL parameters and uncertainties}
    \begin{tabular}{c|c|c|c|c}
        Redshift & $\rm M^{\star}$ & log$_{10}(\phi^{\star})$ & $\alpha$ & $\beta$ \\ \hline
        5 & -21.52$_{-0.37}^{+0.39}$ & -3.52$_{-0.25}^{+0.25}$ & -1.90$_{-0.08}^{+0.08}$ & -5.37$_{-0.38}^{+0.32}$\\
        6 & -21.03$_{-0.37}^{+0.39}$ & -3.52$_{-0.25}^{+0.25}$ & -1.99$_{-0.08}^{+0.08}$ & -4.92$_{-0.38}^{+0.32}$\\
        7 & -20.39$_{-0.37}^{+0.39}$ & -3.48$_{-0.25}^{+0.25}$ & -2.01$_{-0.08}^{+0.08}$ & -3.89$_{-0.38}^{+0.32}$\\
        8 & -20.42$_{-0.37}^{+0.39}$ & -3.70$_{-0.25}^{+0.25}$ & -2.02$_{-0.08}^{+0.08}$ & -3.96$_{-0.38}^{+0.32}$ \\
        9 & -20.45$_{-0.28}^{+0.25}$ & -3.93$_{-0.18}^{+0.18}$ & -2.04$_{-0.07}^{+0.08}$ & -4.03$_{-0.49}^{+0.53}$ \\
        10 & -20.49$_{-0.37}^{+0.31}$ & -4.15$_{-0.27}^{+0.23}$ & -2.06$_{-0.14}^{+0.15}$ & -4.10$_{-0.48}^{+0.48}$ \\
        11 & -20.52$_{-0.30}^{+0.39}$ & -4.38$_{-0.29}^{+0.29}$ & -2.07$_{-0.14}^{+0.15}$ & -4.17$_{-0.48}^{+0.46}$ \\
        12 & -20.55$_{-0.42}^{+0.40}$ & -4.61$_{-0.31}^{+0.30}$ & -2.09$_{-0.21}^{+0.22}$ & -4.24$_{-0.48}^{+0.47}$ \\
        13 & -20.58$_{-0.42}^{+0.40}$ & -4.83$_{-0.31}^{+0.30}$ & -2.10$_{-0.21}^{+0.22}$ & -4.31$_{-0.48}^{+0.47}$ \\
        14 & -20.61$_{-0.42}^{+0.40}$ & -5.06$_{-0.31}^{+0.30}$ & -2.12$_{-0.21}^{+0.22}$ & -4.38$_{-0.48}^{+0.47}$\\
        15 & -20.65$_{-0.42}^{+0.40}$  & -5.29$_{-0.31}^{+0.30}$ & -2.14$_{-0.21}^{+0.22}$ & -4.45$_{-0.48}^{+0.47}$  \\
    \end{tabular}

    \label{tab:DPL}
\end{table}

\section{Neutral gas fraction}
\begin{table}[h!]
    \centering
    \caption{$x_{\rm HI}$ and log($\dot{n}_{\rm ion})$ evaluated at integer redshifts with uncertainties}
    \begin{tabular}{c|c|c|c|c}
        z & log($\dot{n}_{\rm ion}/ \rm s^{-1} Mpc^{-3} $) & $x_{\rm HI}$ & GLIMPSE & D24\\ \hline
    
         15 & 49.3$_{-0.29}^{+0.35}$ & 1.00 & $\pm$ 0.00 & $\pm$ 0.00\\
         14 & 49.41$_{-0.28}^{+0.34}$ & 0.99 & $\pm$ 0.01 & $\pm$ 0.01\\
         13 &  49.70$_{-0.27}^{+0.33}$ & 0.99 & $\pm$ 0.01 & $_{-0.02}^{+0.01}$ \\
         12 & 49.88$_{-0.27}^{+0.32}$ & 0.98 & $_{-0.02}^{+0.01}$& $_{-0.02}^{+0.01}$\\
         11 & 50.10$_{-0.18}^{+0.20}$ & 0.97 & $_{-0.03}^{+0.02}$& $_{-0.04}^{+0.01}$\\
         10 & 50.27$_{-0.17}^{+0.17}$ & 0.94 & $_{-0.04}^{+0.03}$& $_{-0.09}^{+0.02}$\\
         9 & 50.47$_{-0.10}^{+0.09}$ & 0.88 & $_{-0.06}^{+0.04}$& $_{-0.18}^{+0.02}$\\
         8 & 50.64$_{-0.09}^{+0.09}$ & 0.76 & $_{-0.08}^{+0.06}$& $_{-0.38}^{+0.04}$\\
         7 & 50.83$_{-0.09}^{+0.09}$ & 0.51 & $_{-0.12}^{+0.10}$& $_{-0.51}^{+0.09}$\\
         6 & 51.05$_{-0.09}^{+0.09}$ & 0.00 & $_{-0.00}^{+0.14}$& $_{-0.00}^{+0.26}$\\
         5 & 51.12$_{-0.09}^{+0.09}$ & 0.00 & $\pm$ 0& $\pm $ 0.00
    \end{tabular}
\tablefoot{The evolution of $x_{\rm HI}$ and log($\dot{n}_{\rm ion}$) evaluated at integer redshift. For $x_{\rm HI}$ we also show the uncertainties both for the case of the GLIMPSE LF and for the case of the \cite{Donnan_24} (D24) LF, as also seen in Fig.~\ref{fig:xHI}.}
    \label{tab:xHI}
\end{table}

\end{document}